\documentclass[preprint2]{aastex}

\usepackage{amssymb,amsmath,amsthm}
\usepackage{graphicx}
\usepackage{mathrsfs}
\usepackage{multicol}
\usepackage{booktabs}
\usepackage{multirow}
\usepackage{empheq}
\usepackage[hidelinks]{hyperref}
\usepackage[FIGTOPCAP]{subfigure}
\usepackage{captcont}

\usepackage{natbib}
\shorttitle{MCP VII}
\shortauthors{Pesce et al.}

\begin{document}

\title{The Megamaser Cosmology Project. VII. Investigating disk physics using spectral monitoring observations}

\author{D. W. Pesce\altaffilmark{1}, J. A. Braatz\altaffilmark{2}, J. J. Condon\altaffilmark{2}, F. Gao\altaffilmark{2,3}, C. Henkel\altaffilmark{4,5}, E. Litzinger\altaffilmark{6,7}, K. Y. Lo\altaffilmark{2}, M. J. Reid\altaffilmark{8}}

\altaffiltext{1}{Department of Astronomy, University of Virginia, 530 McCormick Road, Charlottesville, VA 22904, USA}
\altaffiltext{2}{National Radio Astronomy Observatory, 520 Edgemont Road, Charlottesville, VA 22903, USA}
\altaffiltext{3}{Shanghai Astronomical Observatory, Chinese Academy of Sciences, 200030 Shanghai, China}
\altaffiltext{4}{Max-Planck-Institut f\"ur Radioastronomie, Auf dem H\"ugel 69, 53121 Bonn, Germany}
\altaffiltext{5}{Astronomy Department, Faculty of Science, King Abdulaziz University, P.O. Box 80203, Jeddah 21589, Saudi Arabia}
\altaffiltext{6}{
Institut f\"ur Theoretische Physik und Astrophysik, Universit\"at W\"urzburg, Emil-Fischer-Str. 31, 97074 W\"urzburg, Germany}
\altaffiltext{7}{Dr. Remeis Sternwarte \& ECAP, Universit\"at Erlangen-N\"urnberg, Sternwartstrasse 7, 96049 Bamberg, Germany}
\altaffiltext{8}{Harvard-Smithsonian Center for Astrophysics, 60 Garden Street, Cambridge, MA 02138, USA}

%\email{dpesce@virginia.edu}
%\email{jbraatz@nrao.edu}
%\email{jcondon@nrao.edu}
%\email{fgao@nrao.edu}
%\email{chenkel@mpifr-bonn.mpg.de}
%\email{eugenia.litzinger@astro.uni-wuerzburg.de}
%\email{flo@nrao.edu}
%\email{mreid@cfa.harvard.edu}

\begin{abstract}
We use single-dish radio spectra of known 22 GHz H$_2$O megamasers, primarily gathered from the large dataset observed by the Megamaser Cosmology Project, to identify Keplerian accretion disks and to investigate several aspects of the disk physics.  We test a mechanism for maser excitation proposed by \cite{mao98}, whereby population inversion arises in gas behind spiral shocks traveling through the disk.  Though the flux of redshifted features is larger on average than that of blueshifted features, in support of the model, the high-velocity features show none of the predicted systematic velocity drifts.  We find rapid intra-day variability in the maser spectrum of ESO 558-G009 that is likely the result of interstellar scintillation, for which we favor a nearby ($D \approx 70$ pc) scattering screen.  In a search for reverberation in six well-sampled sources, we find that any radially-propagating signal must be contributing $\lesssim$10\% of the total variability.  We also set limits on the magnetic field strengths in seven sources, using strong flaring events to check for the presence of Zeeman splitting.  These limits are typically 200--300 mG ($1\sigma$), but our most stringent limits reach down to 73 mG for the galaxy NGC 1194.
\end{abstract}


\section{Introduction}

The Megamaser Cosmology Project (MCP) aims to determine the value of $H_0$ by measuring angular-diameter distances to galaxies in the Hubble flow.  Using the megamaser technique pioneered on the galaxy NGC 4258 \citep[see][]{her99}, the MCP has published distances to the galaxies UGC 3789 \citep{rei13}, NGC 6264 \citep{kuo13}, and NGC 6323 \citep{kuo15}, and additional galaxies are currently being measured.  The ongoing project is a multi-year effort of surveying, monitoring, and mapping maser disks using the Robert C. Byrd Green Bank Telescope (GBT), the Karl G. Jansky Very Large Array (VLA), and the Very Long Baseline Array (VLBA) plus the 100-meter Effelsberg telescope.

The MCP's monitoring campaign uses the GBT of the National Radio Astronomy Observatory (NRAO) to take regular ($\sim$monthly) spectra of megamaser sources targeted for distance measurements.  These spectra are used to measure the accelerations of maser features as part of the determination of $H_0$.  Here we take advantage of this rich dataset to probe the innermost parsec ($\sim$0.1-0.5 pc) of the AGN.  These size scales are roughly an order of magnitude smaller than the dust structures that have been resolved by optical/infrared interferometric studies of the torus region in nearby AGN (see, e.g., \citealt{jaf04}).

The structure of this paper is as follows.  The observations and data reduction procedures are described in \S\ref{ObservationsAndReduction}.  We present the data in \S\ref{DiskSpectra} and \S\ref{DynamicSpectra}, in the form of time-averaged and dynamic spectra, respectively.  In \S\ref{DiskExcitation} we examine a theory of disk maser excitation proposed by \cite{mao98} (hereafter MM98), in \S\ref{Scintillation} we present evidence for the presence of interstellar scintillation in ESO 558-G009, in \S\ref{DiskReverberation} we check the maser disks for signs of propagating disturbances, and in \S\ref{MagneticFieldLimits} we use the spectra to place limits on the magnetic field strengths in the maser disks.

\section{Observations and data reduction} \label{ObservationsAndReduction}

The analyses presented in this paper are based on 22 GHz water maser spectra, almost all of which were taken using the GBT over the period March 2003 -- April 2015.  The majority of these spectra were obtained as part of the survey and monitoring components of the MCP; see \cite{rei09} and \cite{bra10} for details.  We include several non-MCP spectra from the NRAO data archive, most notably for the galaxies NGC 4258 and NGC 3393.

For each MCP spectrum the GBT spectrometer was configured with two 200 MHz spectral windows, one of which was centered on the recession velocity of the galaxy while the other was offset redward by 180 MHz.  Each window had 8192 channels spaced at 24 kHz channel width, which at 22 GHz corresponds to approximately 0.33 km s$^{-1}$.  Both left circular polarization (LCP) and right circular polarization (RCP) were observed simultaneously in each of the two beams of the K-band receiver, and the telescope was nodded on a 2.5-minute cycle to alternate which beam was pointed at the target.  Observations after May 2011 used two of the seven beams of the K-band Focal Plane Array (KFPA) in the same nodding scheme.  Integration times for the monitored sources were typically between 1 and 3 hours during a single observing session.

We reduced GBT data using the same methods outlined in previous MCP papers (see, e.g., \citealt{bra10}).  Our measurements of Zeeman splitting (\S\ref{ZeemanMeasurements}) use spectra at their native resolution, prior to Hanning smoothing.

Integrated line fluxes in some of our spectra are affected by a broad ($\sim$1500 km s$^{-1}$) sinusoidal baseline ripple.  The baseline ripples are generally comparable in amplitude to the RMS channel noise, but their contributions to the flux measurements can be the dominant source of uncertainty for our best-sampled sources.  To characterize the flux uncertainty from the baseline ripple, we averaged the frequency-offset spectral windows from each observation.  These spectra are free of maser emission, were taken concurrently with the science spectra using the same instrument configuration on the GBT, and have all undergone the same data reduction procedure.  We measured the RMS of the integrated flux inside a boxcar window placed randomly inside the averaged spectrum, as a function of the spectral width of that window.  The line flux uncertainty behaves approximately quadratically as a function of window width, reaching a maximum of $\sim$0.1 Jy km s$^{-1}$ for a window width of $\sim$750 km s$^{-1}$.  We thus assign a baseline ripple uncertainty to each line flux measurement that follows the empirical relation given by

\begin{equation}
\sigma_{\text{S},0.1} = - \left( \Delta v \right)_{750}^2 + 2 \left( \Delta v \right)_{750} \label{eqn:BaselineUncertainty}
\end{equation}

\noindent Here, $\sigma_{\text{S},0.1}$ is the baseline rippler's contribution to the line flux uncertainty (in units of 0.1 Jy km s$^{-1}$) and $(\Delta v)_{750}$ is the spectral window width (in units of 750 km s$^{-1}$).

\section{Identifying Keplerian disk megamasers} \label{DiskSpectra}

Our aim is to examine the spectral characteristics of maser emission from accretion disks, including flux ratios, secular velocity drifts, and variability.  We thus seek to identify maser systems with spectra dominated by emission from edge-on, Keplerian disks.

To date, 16 megamaser disk systems have published VLBI maps.  Eight of these were mapped by the MCP (NGC 1194, NGC 2273, Mrk 1419, NGC 4388, NGC 6323 in \citealt{kuo11}; UGC 3789 in \citealt{rei09}; NGC 6264 in \citealt{kuo13}; and NGC 5765b in Gao et al. \textit{submitted}), and eight were mapped by other groups (NGC 1068 in \citealt{gre97b}; NGC 4945 in \citealt{gre97c}; NGC 5793 in \citealt{hag01}; Circinus in \citealt{gre03b}; NGC 3079 in \citealt{kon05}; NGC 4258 in \citealt{arg07}; NGC 3393 in \citealt{kon08};  and IC 1481 in \citealt{mam09}).  Of these 16 mapped disks, nine have ``clean" Keplerian rotation curves, and all nine share a distinctive single-dish spectral profile.  To maximize the uniformity and size of the sample for the analysis in this paper, we therefore selected sources based on the appearance of their single-dish (usually GBT) spectra.

A ``clean" disk megamaser is an edge-on maser in Keplerian rotation around the central SMBH, in which the disk maser emission dominates over any jet or outflow maser components.  These systems have characteristic spectra that are marked by three sets of maser components.  The ``systemic" set of features coincides roughly with the recession velocity of the galaxy, and the masing arises along a line of sight through the disk to the central AGN.  The two ``high-velocity" sets of features (the ``redshifted features" and ``blueshifted features") are spectrally offset to either side of the galaxy's recession velocity.  These features arise from the midline of the accretion disk, along lines of sight that are tangent to the orbital motion (which ensures velocity coherence throughout the column of gas).  For an edge-on disk, the midline is the diameter through the disk that falls perpendicular to the line of sight.

To select clean disk megamasers, we use the following criteria.  The spectra must show at least two of the three expected distinct sets of maser features (in maser disks with only two sets of features, the third set is presumably present but below the detection threshold).  Furthermore, at least one of the sets of features should have components that are offset from the recession velocity by at least 300 km s$^{-1}$ (an empirically-determined high galactic rotation cutoff; see \citealt{cre09}), to avoid contaminating the sample with interstellar masers (from, e.g., a strong starburst) and sub-Keplerian rotators.  For a spectrum with only two sets of maser features, we require either that one of these feature sets be coincident with the recession velocity of the galaxy or that both feature sets be offset from the recession velocity by at least 300 km s$^{-1}$.

Though we have attempted to be comprehensive in our selection of sources, there are several known disk (or disk-like) H$_2$O megamasers that do not make it into our sample because the disk emission is contaminated by non-disk components.  Circinus \citep{gar82} contains a masing accretion disk, but it also has maser emission associated with an outflow \citep{gre03b}.  Similarly, NGC 1068 \citep{cla84} has maser emission arising from both a disk and a radio jet encountering a dense molecular cloud \citep{gal96}.  Complexities like these confuse the maser spectrum and make it difficult to associate individual spectral features with either the disk or outflow/jet components without a VLBI map.  For this reason, none of these sources passes our selection criteria.

\begin{figure*}[p]
	\centering
		\caption{Spectra for disk megamasers used in our analysis of the MM98 model.  Each spectrum is a weighted average (see \S\ref{DiskSpectra}) taken over all epochs; the date of the first epoch is located at the top right.  Galaxy recession velocities and associated 1$\sigma$ errors (see Table \ref{tab:Maoz_McKee_targets}) are overplotted in red.} \label{fig:AveragedSpectra}
		\subfigure{\includegraphics[width=1.0\textwidth]{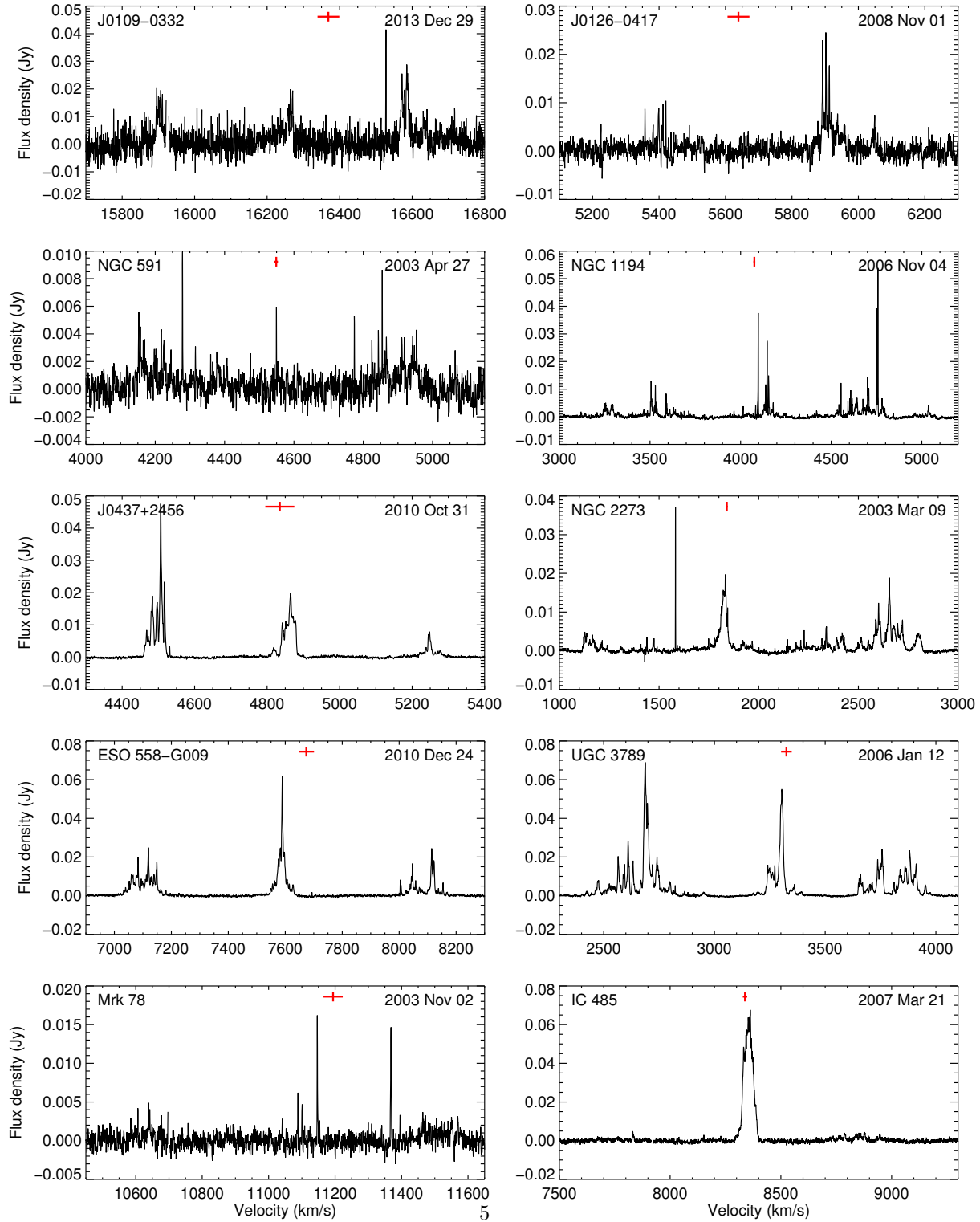}}
\end{figure*}

The final list of 32 clean megamaser disk systems used in our study is given in Table \ref{tab:Maoz_McKee_targets}.  Figure \ref{fig:AveragedSpectra} shows the weighted average spectra of these sources, where the weighting $\tau / T_{\text{sys}}^2$ (where $\tau$ is the exposure time and $T_{\text{sys}}$ is the system temperature) was chosen to minimize the RMS noise of each spectrum.  The emission from the remaining $\sim$130 known water megamaser galaxies may arise from nuclear sources other than the accretion disk (e.g., molecular gas in an outflow) or from extranuclear sources elsewhere within these galaxies (e.g., star-forming regions).

For completeness, we reproduce the spectrum of ESO 269-G012 in Figure \ref{fig:AveragedSpectra} from \cite{gre03a}; see that paper for details about the observation and data reduction.

\begin{deluxetable}{lcccccccccccc}
\rotate
\tabletypesize{\tiny}
\tablecolumns{9}
\tablewidth{0pt}
\tablecaption{\label{tab:Maoz_McKee_targets}} 
\tablehead{	&	\colhead{R.A.}	&	\colhead{Dec.}	&	\colhead{$V_{\text{rec}}$}	&	\colhead{Velocity}	&	\colhead{$\tau$} &	\colhead{RMS}	&	\colhead{$L_{\text{iso}}$}	&	\colhead{Blue}	&	\colhead{Sys}	&	\colhead{Red}	&		& \\
\cmidrule[0.5pt](lr){9-11}
Target &	\colhead{(J2000)}	&	\colhead{(J2000)}	&	\colhead{(km s$^{-1}$)}	&	\colhead{type}	&	\colhead{(hours)}	&	\colhead{(mJy)}	&	\colhead{($L_{\odot} h_{70}^{-2}$)}	&	\multicolumn{3}{c}{(Jy km s$^{-1}$)}	&	\colhead{$\log(R)$}	&	\colhead{Ref.}}
\startdata
J0109-0332		&	01:09:45.1	&	$-03$:32:33	&	$16369 \pm 30$												&	O	&	\hphantom{10}1.7	&	3.08	&	$2086 \pm 180$	&	0.37	&	0.48			&	0.86	&	$0.37 \pm 0.10$			&	(a)	\\
J0126-0417		&	01:26:01.7	&	$-04$:17:56	&	\hphantom{1}$5639 \pm 33$							&	O	&	\hphantom{10}3.2	&	1.40	&	$105 \pm 18$		&	0.11	&	$< 0.08$	&	0.48	&	$0.64 \pm 0.16$			&	(a)	\\
NGC 591				&	01:33:31.2	&	$+35$:40:06	&	\hphantom{1}$4549 \pm 5$\hphantom{0}	&	H	&	\hphantom{10}6.6	&	0.69	&	\hphantom{0}$38 \pm 10$			&	0.20	&	0.01	&	0.16	&	$-0.10 \pm 0.41$\hphantom{$-$}		&	(b)	\\
NGC 1194			&	03:03:49.1	&	$-01$:06:13	&	\hphantom{1}$4076 \pm 5$\hphantom{0}	&	H	&	100.8							&	0.24	&	$131 \pm 7$\hphantom{0}			&	0.51	&	0.38			&	0.79	&	$0.19 \pm 0.12$			&	(a)	\\
J0437+2456		&	04:37:03.7	&	$+24$:56:07	&	\hphantom{1}$4835 \pm 40$							&	O	&	119.2							&	0.25	&	$155 \pm 11$		&	0.70	&	0.53			&	0.16	&	$-0.64 \pm 0.16$\hphantom{$-$}			&	(a)	\\
NGC 2273			&	06:50:08.7	&	$+60$:50:45	&	\hphantom{1}$1840 \pm 4$\hphantom{0}	&	H	&	\hphantom{0}99.5	&	0.27	&	$37 \pm 1$			&	0.25	&	0.73			&	1.33	&	$0.73 \pm 0.20$			&	(e)	\\
ESO 558-G009	&	07:04:21.0	&	$-21$:35:19	&	\hphantom{1}$7674 \pm 27$							&	O	&	114.0							&	0.29	&	$709 \pm 14$		&	0.96	&	0.98			&	0.63	&	$-0.18 \pm 0.06$\hphantom{$-$}				&	(a)	\\
UGC 3789			&	07:19:31.6	&	$+59$:21:21	&	\hphantom{1}$3325 \pm 24$							&	H	&	187.8							&	0.16	&	$357 \pm 2$\hphantom{0}			&	3.17	&	1.73			&	2.00	&	$-0.20 \pm 0.02$\hphantom{$-$}				&	(f)	\\
Mrk 78				&	07:42:41.7	&	$+65$:10:37	&	$11194 \pm 29$												&	O	&	\hphantom{10}4.3	&	0.87	&	$104 \pm 60$	&	0.05	&	0.04			&	0.14	&	$0.45 \pm 1.25$			&	(b)	\\
IC 485				&	08:00:19.8	&	$+26$:42:05	&	\hphantom{1}$8338 \pm 10$							&	H	&	\hphantom{10}8.0	&	0.74	&	$1061 \pm 26$\hphantom{0}	&	0.03	&	3.06			&	0.29	&	$0.99 \pm 0.18$			&	(a)	\\
J0836+3327		&	08:36:22.8	&	$+33$:27:39	&	\hphantom{0}$14810 \pm 120$						&	O	&	\hphantom{10}2.4	&	1.16	&	$937 \pm 67$	&	0.24	&	0.55	&	0.17	&	$-0.15 \pm 0.21$\hphantom{$-$}				&	(g)	\\
J0847-0022		&	08:47:47.7	&	$-00$:22:51	&	$15275 \pm 32$												&	O	&	\hphantom{10}1.4	&	2.37	&	$2945 \pm 129$	&	0.83	&	0.62			&	1.29	&	$0.19 \pm 0.06$										&	(a)	\\
Mrk 1419			&	09:40:36.4	&	$+03$:34:37	&	\hphantom{1}$4947 \pm 7$\hphantom{0}	&	H	&	151.4							&	0.20	&	$565 \pm 5$\hphantom{0}	&	2.42	&	1.07			&	1.39	&	$-0.24 \pm 0.03$\hphantom{$-$}				&	(h)	\\
IC 2560				&	10:16:18.7	&	$-33$:33:50	&	\hphantom{1}$2925 \pm 2$\hphantom{0}	&	H	&	\hphantom{0}27.4	&	0.79	&	$210 \pm 4$\hphantom{0}	&	0.71	&	3.55			&	1.07	&	$0.18 \pm 0.05$			&	(d)	\\
Mrk 34				&	10:34:08.6	&	$+60$:01:52	&	$15292 \pm 12$												&	O	&	\hphantom{10}3.8	&	0.41	&	$814 \pm 64$	&	0.52	&	$< 0.11$	&	0.37	&	$-0.15 \pm 0.14$\hphantom{$-$}										&	(i)	\\
NGC 3393			&	10:48:23.4	&	$-25$:09:43	&	\hphantom{1}$3750 \pm 5$\hphantom{0}	&	H	&	\hphantom{10}5.0	&	0.66	&	$259 \pm 3$\hphantom{0}		&	1.39	&	0.74			&	1.91			&	$0.14 \pm 0.03$			&	(g)	\\
UGC 6093			&	11:00:48.0	&	$+10$:43:41	&	$10805 \pm 10$												&	H	&	\hphantom{0}40.5	&	0.33	&	$1048 \pm 24$\hphantom{0}	&	0.43	&	0.87			&	0.63	&	$0.17 \pm 0.07$			&	(a)	\\
NGC 4258			&	12:18:57.5	&	$+47$:18:14	&	\hphantom{10}$448 \pm 3$\hphantom{0}	&	H	&	\hphantom{0}21.6	&	0.66	&	\hphantom{$^{\dag}$}$89.7 \pm 0.2^{\dag}$	&	0.36	&	57.0			&	9.45	&	$1.42 \pm 0.09$			&	(j)	\\
NGC 4388			&	12:25:46.7	&	$+12$:39:44	&	\hphantom{1}$2517 \pm 4$\hphantom{0}	&	H	&	\hphantom{0}13.4	&	0.63	&	$13 \pm 3$	&	0.15	&	$< 0.05$		&	0.30	&	$0.30 \pm 0.41$			&	(b)	\\
ESO 269-G012	&	12:56:40.5	&	$-46$:55:34	&	\hphantom{1}$5014 \pm 13$							&	H	&	\hphantom{10}1.4	&	10.2	&	$496 \pm 53$	&	2.68	&	0.10	&	1.90	&	$-0.15 \pm 0.04$\hphantom{$-$}				&	(k)	\\
NGC 4968			&	13:07:06.0	&	$-23$:40:37	&	\hphantom{1}$2988 \pm 15$							&	O	&	\hphantom{10}3.8	&	2.92	&	$54 \pm 5$	&	0.08	&	0.53			&	0.62	&	$0.89 \pm 0.21$			&	(a)	\\
J1346+5228		&	13:46:40.8	&	$+52$:28:37	&	\hphantom{1}$8737 \pm 12$							&	O	&	\hphantom{0}10.6	&	0.68	&	$380 \pm 29$	&	0.18	&	0.84			&	0.16	&	$-0.05 \pm 0.35$\hphantom{$-$}				&	(a)	\\
CGCG 074-064	&	14:03:04.5	&	$+08$:56:51	&	\hphantom{1}$6886 \pm 29$							&	O	&	\hphantom{10}5.3	&	1.62	&	$852 \pm 21$	&	0.44							&	2.78	&	0.56			&	$0.49 \pm 0.06$					&	(a)	\\
NGC 5495			&	14:12:23.3	&	$-27$:06:29	&	\hphantom{1}$6737 \pm 9$\hphantom{0}	&	H	&	\hphantom{10}6.7	&	0.85	&	$625 \pm 18$	&	0.47	&	1.93			&	0.53	&	$-0.05 \pm 0.10$\hphantom{$-$}										&	(g)	\\
NGC 5765b			&	14:50:51.5	&	$+05$:06:52	&	\hphantom{1}$8333 \pm 19$							&	O	&	\hphantom{0}62.2	&	0.37	&	$2553 \pm 17$\hphantom{0}	&	1.10	&	5.56			&	1.11	&	$0.00 \pm 0.05$			&	(a)	\\
UGC 9618b			&	14:57:00.7	&	$+24$:37:03	&	$10094 \pm 5$\hphantom{0}							&	O	&	\hphantom{10}3.5	&	0.97	&	$794 \pm 54$	&	0.19	&	0.54	&	0.90	&	$0.68 \pm 0.39$			&	(g)	\\
UGC 9639			&	14:58:36.0	&	$+44$:53:01	&	$10886 \pm 17$												&	O	&	\hphantom{10}8.3	&	0.88	&	$264 \pm 62$	&	0.16	&	0.10			&	0.23	&	$0.16 \pm 0.29$			&	(a)	\\
CGCG 165-035	&	15:14:39.8	&	$+$26:35:39	&	\hphantom{1}$9622 \pm 2$\hphantom{0}	&	H	&	\hphantom{0}0.85	&	1.68	&	$1447 \pm 22$\hphantom{0}	&	1.71				&	0.75	&	0.88			&	$-0.29 \pm 0.05$\hphantom{$-$}			&	(a)	\\
NGC 6264			&	16:57:16.1	&	$+27$:50:59	&	$10177 \pm 28$												&	O	&	103.5							&	0.23	&	$1634 \pm 11$\hphantom{0}	&	1.07	&	0.81			&	1.48	&	$0.14 \pm 0.05$			&	(g)	\\
J1658+3923		&	16:58:15.5	&	$+39$:23:29	&	$10292 \pm 11$												&	O	&	\hphantom{10}3.2	&	1.41	&	$427 \pm 44$	&	0.41	&	$< 0.12$	&	0.65	&	$0.20 \pm 0.10$			&	(a)	\\
NGC 6323			&	17:13:18.0	&	$+43$:46:56	&	\hphantom{1}$7772 \pm 35$							&	O	&	134.1							&	0.19	&	$839 \pm 21$	&	0.85	&	0.45			&	1.65	&	$0.29 \pm 0.03$			&	(b)	\\
CGCG 498-038	&	23:55:44.2	&	$+30$:12:44	&	\hphantom{1}$9240 \pm 27$							&	O	&	\hphantom{10}2.5	&	1.40	&	$280 \pm 32$	&	0.25	&	0.19			&	0.23	&	$-0.04 \pm 0.42$\hphantom{$-$}										&	(a)	\\
\enddata \vspace{-5mm}
\tablecomments{Observational properties of the 32 disk megamasers.  The recession velocities ($V_{\text{rec}}$) use the optical convention in the heliocentric reference frame (velocities and errors were taken from NED and references therein).  In the ``velocity type" column, H indicates that the recession velocity was measured using HI 21-cm data, O means that it was measured using optical/IR lines.  The total integration time ($\tau$) and final RMS are listed for the averaged spectra (see Figure \ref{fig:AveragedSpectra}).  The isotropic luminosities ($L_{\text{iso}}$) have been calculated assuming a Hubble constant of 70 km s$^{-1}$ Mpc$^{-1}$; the associated uncertainties are statistical and do not account for any systematic flux calibration offsets (which may be as large as $\sim$20\%) or peculiar velocities (which may be important for nearby galaxies).  Columns labeled ``Blue," ``Sys," and ``Red" list the integrated line fluxes for the blueshifted, systemic, and redshifted feature sets, respectively; when measured line fluxes are smaller than the uncertainty, 1$\sigma$ upper limits are listed instead.  The logarithms of the red-to-blue flux ratios are given in the column labeled $\log(R)$.  The reference (``Ref.") column lists a citation for the discovery paper for each source, specified below.  A comprehensive list of extragalactic H$_2$O masers is maintained on the MCP website (\url{https://safe.nrao.edu/wiki/bin/view/Main/PublicWaterMaserList}).  References: (a) Braatz et al. (\textit{in prep.}); (b) \cite{bra04}; (c) \cite{gre09}; (d) \cite{bra96}; (e) \cite{zha06}; (f) \cite{bra08}; (g) \cite{kon03}; (h) \cite{hen02}; (i) \cite{hen05}; (j) \cite{cla84}; (k) \cite{gre03a}. \\ \vspace{1mm}
$^{\dag}$For NGC 4258, the distance measurement from \cite{hum13} of 7.6 Mpc was used instead of the Hubble law value.}
\end{deluxetable}

\begin{figure*}[p]
	\centering
		\captcont{(continued)}
		\subfigure{\includegraphics[width=1.0\textwidth]{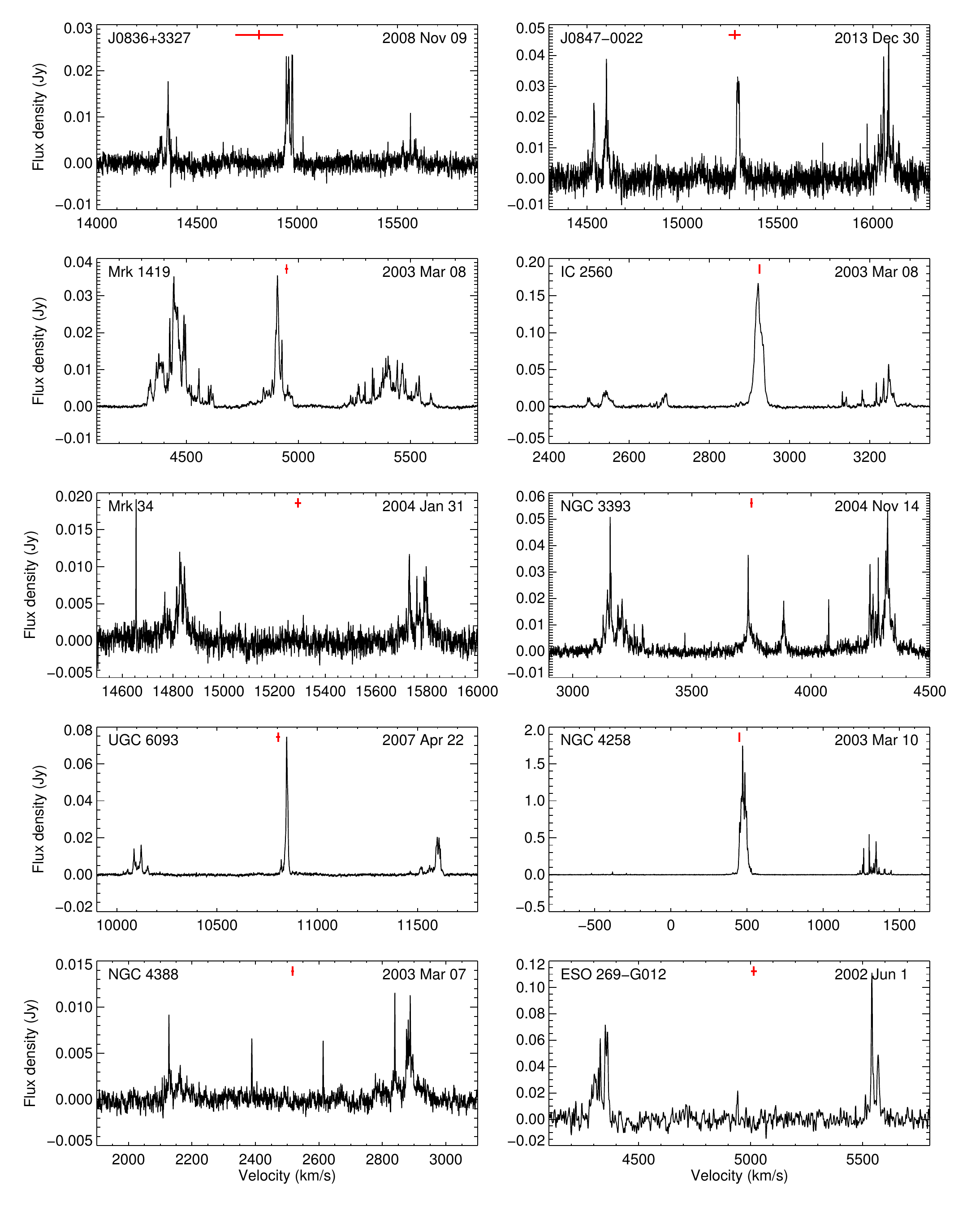}}
\end{figure*}
\begin{figure*}[p]
	\centering
		\captcont{(continued)}
		\subfigure{\includegraphics[width=1.0\textwidth]{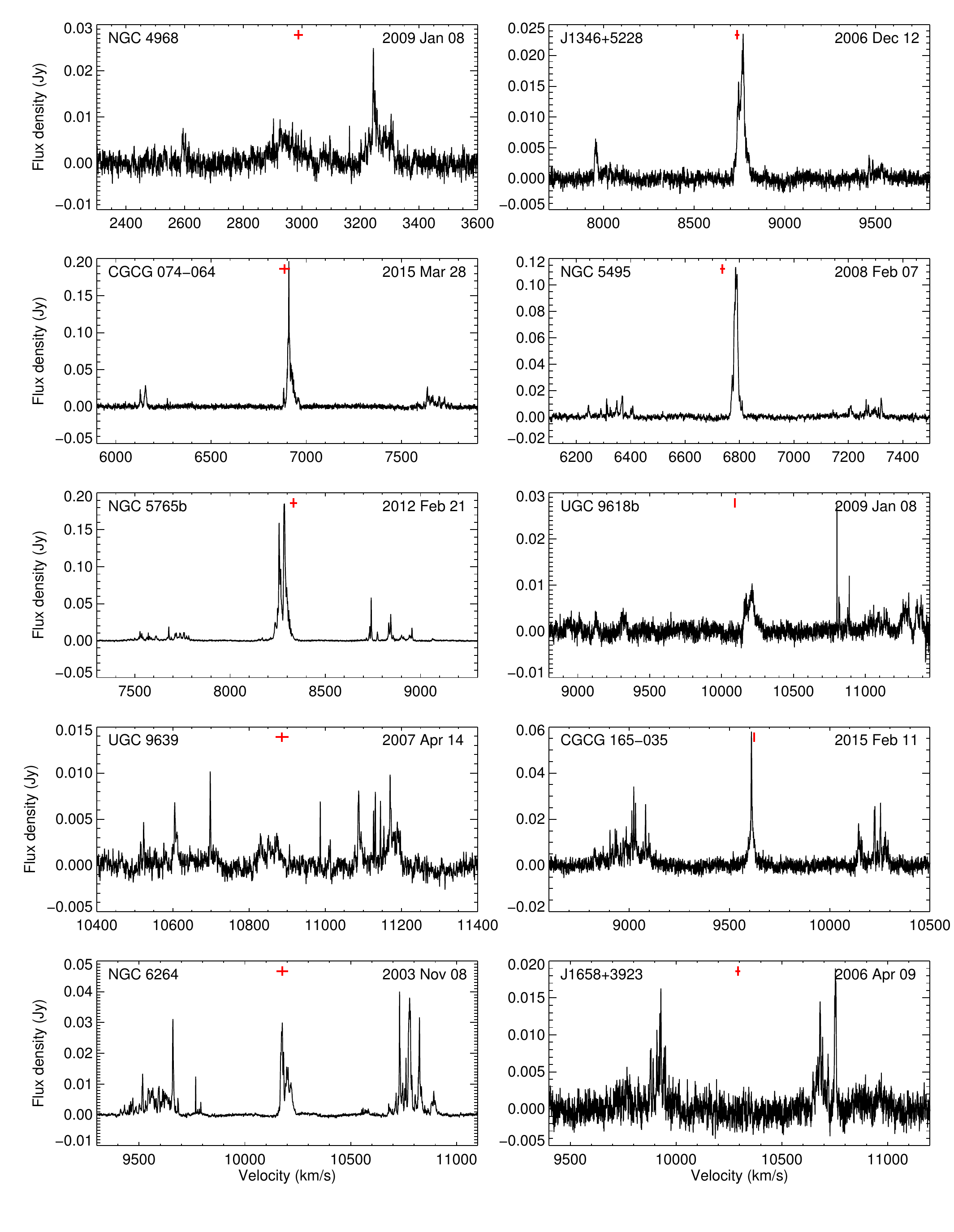}}
\end{figure*}
\begin{figure*}[p]
	\centering
		\captcont{(continued)}
		\subfigure{\includegraphics[width=1.0\textwidth]{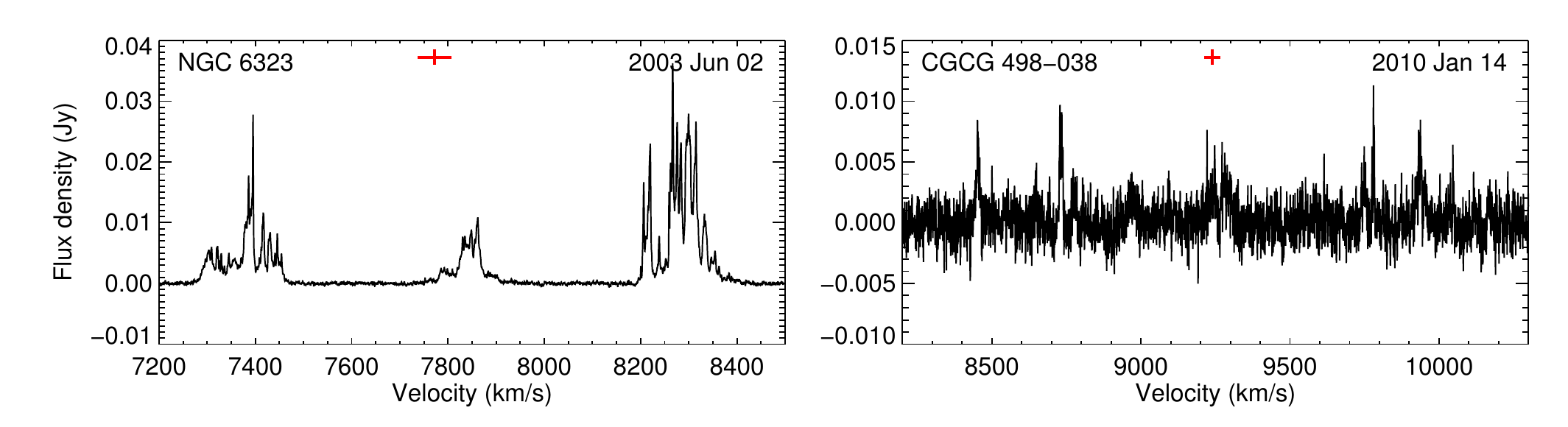}}
		\vspace{50mm}
\end{figure*}

\subsection{Observed properties of disk megamasers} \label{ObservedProperties}

Table \ref{tab:Maoz_McKee_targets} also lists several observational properties of each galaxy.  We obtained the recession velocities from the NASA/IPAC Extragalactic Database (NED), favoring velocities measured from neutral hydrogen (HI) over those made from optical lines.  HI measurements have the advantage that they average over all internal motions of a galaxy, while optical lines are preferentially emitted from regions with a sufficiently energetic radiation field to excite the transitions.  In the case of active galaxies, like those present in our sample, the optical emission could very well be dominated by gas that is kinematically driven by the nuclear activity (e.g., outflows).  This could result in a systematic offset between the recession velocity of the galaxy and the velocity measured using optical lines (e.g., \citealt{com13}).  We do see such offsets in several of the spectra shown in Figure \ref{fig:AveragedSpectra}.

We measured line fluxes separately for each set of features: blueshifted, systemic, and redshifted.  To maximize the signal-to-noise for those spectra with weak features, we integrated only over spectral windows that contained clear signal.  In some cases this meant integrating over several distinct, narrow windows to obtain the total line flux for that set of features.  For several spectra, the systemic set of features is absent; in these cases we list an upper limit on the line flux for the systemic features obtained by integrating over the spectral region located between the high-velocity features (i.e., the region redward of the blueshifted features and blueward of the redshifted features).

To obtain the total isotropic luminosities listed in Table \ref{tab:Maoz_McKee_targets}, we integrated each spectrum across the full span of maser emission.  For a measured line flux $S$, the isotropic luminosity is given by

\begin{equation}
L_{\text{iso}} = \frac{4 \pi v^2 S}{H_0^2}. \label{eqn:IsotropicLuminosity}
\end{equation}

\begin{figure*}
	\centering
		\includegraphics[width=1.0\textwidth]{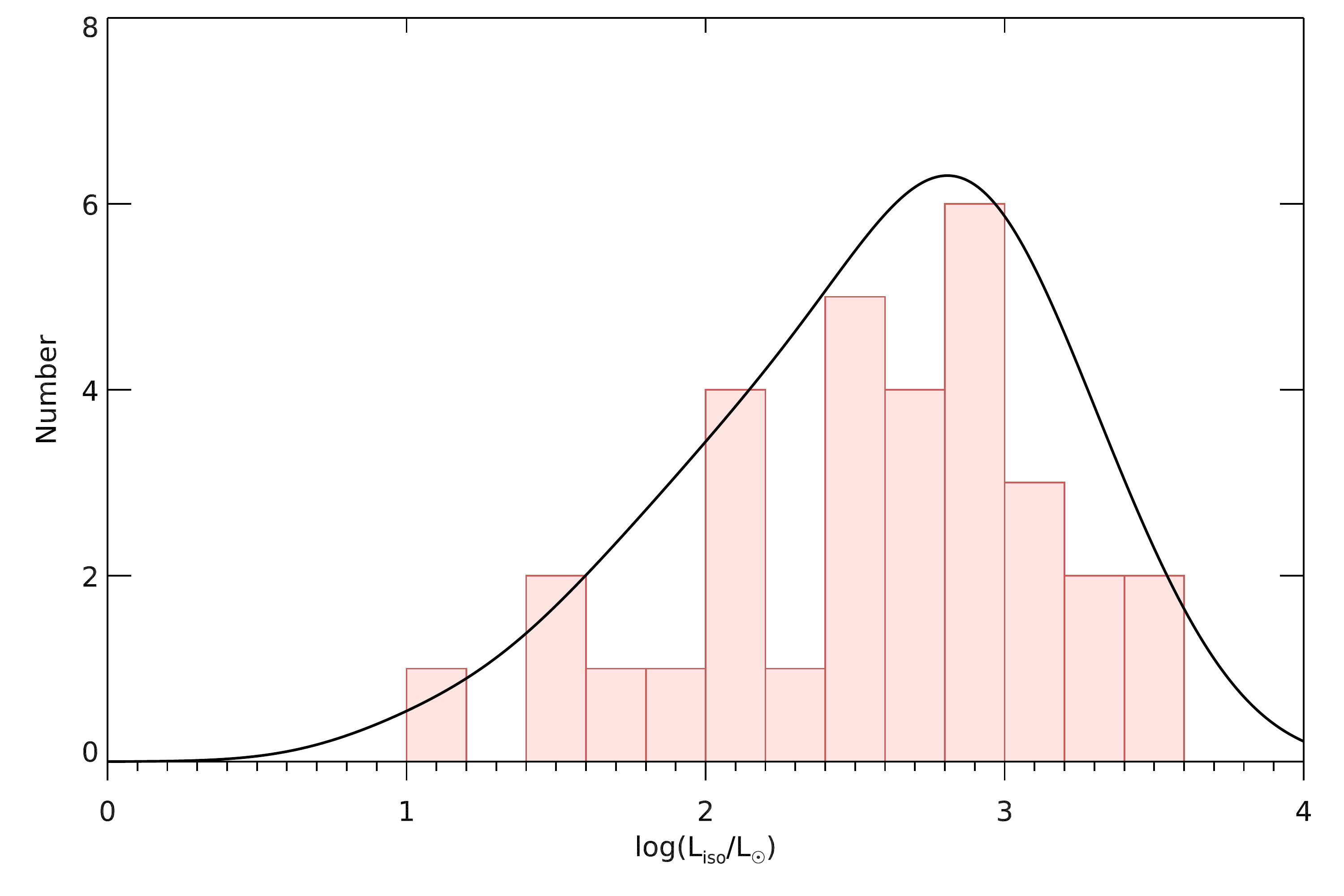}
	\caption{Histogram showing the distribution of isotropic luminosities for our sample of disk megamasers.  The solid black line shows the kernel density estimation obtained using a normal kernel, with Silverman's rule applied for the kernel Gaussian width (see Appendix \ref{app:KDE}).}
	\label{fig:histogram_luminosities}
\end{figure*}

\noindent Here, $v$ is the recession velocity of the galaxy.  This expression is accurate for low-redshift ($z \lesssim 0.1$) sources, and all of our galaxies fall into this category so we use it throughout.  In our calculations, we assume a Hubble constant of $H_0 = 70$ km s$^{-1}$ Mpc$^{-1}$.  Figure \ref{fig:histogram_luminosities} shows a histogram of the isotropic luminosities.  To alleviate the somewhat arbitrary nature of histogram bin sizes and endpoints, we have overplotted a kernel density estimate using a Gaussian kernel.  The area of each kernel is equal to that of a histogram bin with a bin size determined using Silverman's rule; see Appendix \ref{app:KDE} for details.

The measured isotropic luminosities span over two orders of magnitude, and the observed distribution (see Figure \ref{fig:histogram_luminosities}) appears to be consistent with a sensitivity-limited sample (i.e, the highest luminosity masers tend to be found at large distances, and vice versa).  While some of this spread is undoubtedly caused by intrinsic power differences among the many systems, most of it is likely the result of viewing angles.  Though the exact angular dependence of the maser emission is a strong function of the source geometry and saturation, it always drops off exponentially from the beam center, which falls along the path of maximum gain.  (For an in-depth discussion of maser beaming, we refer the interested reader to \citealt{eli92}.)  Thus, even a slight ($\lesssim$5$^{\circ}$) inclination of the maser beam from the line of sight could cause the observed intensity to drop by an order of magnitude or more.  This is especially true if the masers are unsaturated.  The unknown contribution from maser beaming precludes us from correcting the Malmquist bias and turning Figure \ref{fig:histogram_luminosities} into a true luminosity function.

\section{Testing a model of disk maser excitation} \label{DiskExcitation}

In their 1998 paper, \citeauthor{mao98} (MM98) sought to explain the observation in NGC 4258 that the line flux of the redshifted features is much higher than the line flux of the blueshifted features.  In their model, population inversion (and thus masing) only occurs in post-shock gas on the trailing edge of a spiral shock in the accretion disk.  Observed high-velocity maser features then occur wherever the line of sight falls tangent to a shock front, for an edge-on disk system.

The geometry of the trailing spiral shocks causes redshifted maser emission to preferentially originate from the region of the disk that lies in front of the midline, while blueshifted maser emission arises from behind the midline.  The blueshifted photons would thus pass through a sightline of velocity-coherent (but noninverted) gas, leading to absorption that is not present for the redshifted photons.  The model thereby predicts that the redshifted high-velocity features observed for disk maser systems should be systematically stronger than the blueshifted high-velocity features.  See Fig. 1 in MM98 for an illustration of this geometry.

Owing to their offsets from the midline, the MM98 model predicts nonzero line-of-sight ``accelerations"; specifically, the blueshifted features should show a mean positive acceleration while the redshifted features show a negative one.  These arise because as the trailing spiral shock passes through the disk, the inversion region (and thus the segment of spiral structure that is tangent to the line of sight) moves radially outwards with time.  The line-of-sight component of the velocity decreases in magnitude with increasing radius, so the result is an observed velocity drift in the high-velocity maser lines.  Though such behavior mimics an acceleration, it is actually tracing the rotating spiral structure rather than the Keplerian motion of the gas in the disk, and we therefore refer to the phenomenon as a ``velocity drift" rather than as an acceleration (see \S\ref{ExcitationAccelerations} for details).  This prediction runs counter to that of the ``standard" model, which has an entirely masing disk with high-velocity features falling close to the midline.  The standard model predicts that the high-velocity features should have nearly zero line-of-sight accelerations on average.

The model proposed by MM98 was inspired by the red-blue flux asymmetry in NGC 4258, which we note from Table \ref{tab:Maoz_McKee_targets} has a uniquely high value of $\log(R) = +1.42$ not seen in any other maser disk.  It is an open question whether such an excitation mechanism applies to maser disks in general; indeed, it is an open question whether this mechanism even holds for NGC 4258 (see, e.g., \citealt{bra00}).  We checked this model by measuring the flux asymmetry and velocity drifts of high-velocity features in our Keplerian disk sample.

\subsection{Statistical analysis} \label{ExcitationAnalysis}

\begin{figure*}
	\centering
		\includegraphics[width=1.0\textwidth]{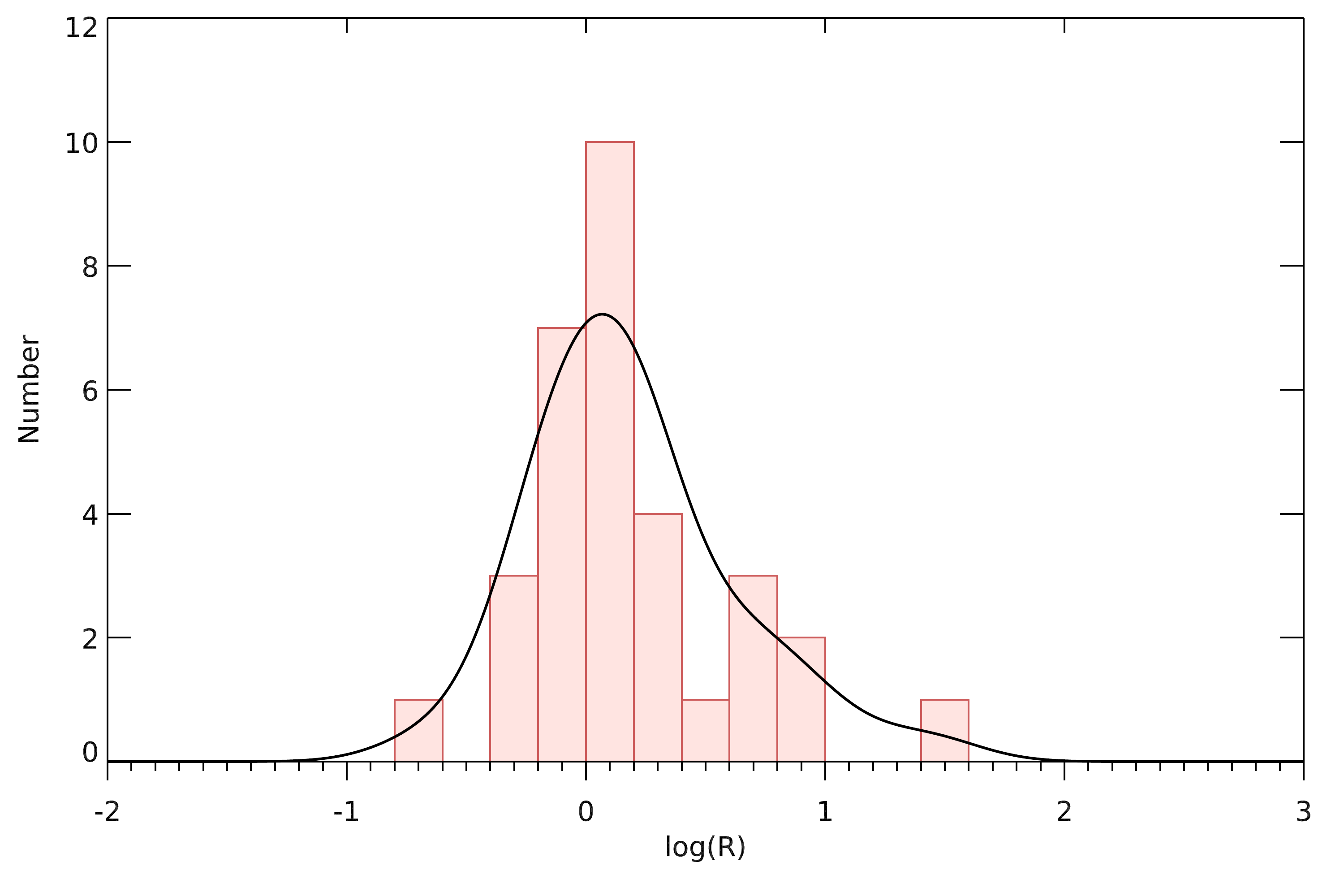}
	\caption{Histogram showing the distribution of the logarithm of the red/blue flux ratios for our sample of disk megamasers.  The solid black line again shows the kernel density estimation, obtained using the same normalization as in Figure \ref{fig:histogram_luminosities}.  NGC 4258 occupies the rightmost histogram bin, causing the red tail of the distribution to be noticeably longer and heavier than the blue tail.  Though we include it in this plot, NGC 4258 was not included in the statistical analysis performed in \S\ref{ExcitationAnalysis} to avoid biasing the results (i.e., since the proposed hypothesis was based on observations of NGC 4258, its observed properties necessarily agree with the hypothesis).}
	\label{fig:histogram_MaozMcKee}
\end{figure*}

For each disk maser in our sample we made a weighted average spectrum from all epochs of observation (see Figure \ref{fig:AveragedSpectra} and \S\ref{DiskSpectra}).  The averaging reduces the noise and mitigates the effects of variability.  We then identified the regions of each spectrum corresponding to the redshifted and blueshifted high-velocity features.  By integrating over these spectral segments, we obtained the redshifted and blueshifted fluxes.  The ratio, $R$, of the redshifted to the blueshifted flux should be greater than 1 for the MM98 model.  The values of $\log(R)$ for our sample are listed in Table \ref{tab:Maoz_McKee_targets} and their histogram is plotted in Figure \ref{fig:histogram_MaozMcKee}.

The null hypothesis is that the redshifted and blueshifted fluxes are on average equal; that is, the logarithm of the ratio of the redshifted to the blueshifted flux should be a distribution centered on zero.  We use the logarithm of the flux ratios (rather than the ratios themselves) to avoid the skewing of the distribution that arises from a direct ratio.

To test whether our results are consistent with the null hypothesis, we employ a likelihood analysis to determine whether the sample we observe has been drawn from a parent population with an intrinsic flux ratio distribution centered on zero.  The data point corresponding to NGC 4258 is not included in this analysis, as it was used to generate the original hypothesis.  Here we utilize a technique analogous to that presented in \cite{ric11}.

To simplify notation, we define $X \equiv \log(R)$, where $R = \rho/\beta$ is the ratio of the redshifted flux (denoted $\rho$) to the blueshifted flux (denoted $\beta$).  We assume that the parent distribution of $X$ is a Gaussian centered on $X_0$, with a standard deviation of $\sigma_0$.  We also assume that the observational uncertainties associated with each measurement are normally distributed about the intrinsic value for that measurement.

For a single observation of a source with intrinsic redshifted flux of $\rho_t$, the probability to observe the value $\rho_i$ with uncertainty $\sigma_{r,i}$ is given by

\begin{equation}
P_r = \frac{1}{\sigma_{r,i} \sqrt{2 \pi}} \exp\left[ - \frac{(\rho_t - \rho_i)^2}{2 \sigma_{r,i}^2} \right]. \label{eqn:Likelihood1_red}
\end{equation}

\noindent Similarly for an observation of a source with intrinsic blueshifted flux of $\beta_t$, the probability to observe the value $\beta_i$ with uncertainty $\sigma_{b,i}$ will be

\begin{equation}
P_b = \frac{1}{\sigma_{b,i} \sqrt{2 \pi}} \exp\left[ - \frac{(\beta_t - \beta_i)^2}{2 \sigma_{b,i}^2} \right]. \label{eqn:Likelihood1_blue}
\end{equation}

We also have the probability for the source to have an intrinsic flux ratio of $X_t = \log(\rho_t/\beta_t)$, given the parent distribution

\begin{equation}
P_t = \frac{1}{\sigma_0 \sqrt{2 \pi}} \exp\left[ - \frac{(X_t - X_0)^2}{2 \sigma_0^2} \right]. \label{eqn:Likelihood2}
\end{equation}

\noindent The resulting likelihood of the observation is then given by an integral over the product of these probability density functions,

\begin{equation}
\ell_i = \int_0^{\infty} \int_0^{\infty} P_r P_b P_t d\rho_t d\beta_t . \label{eqn:Likelihood3}
\end{equation}

\noindent For $N$ observations, the joint likelihood will then be the product of the individual measurement likelihoods:

\begin{equation}
\mathcal{L}(X_0,\sigma_0) = \prod_{i=1}^N \ell_i . \label{eqn:Likelihood5}
\end{equation}

\begin{figure}[t]
		\includegraphics[width=0.48\textwidth]{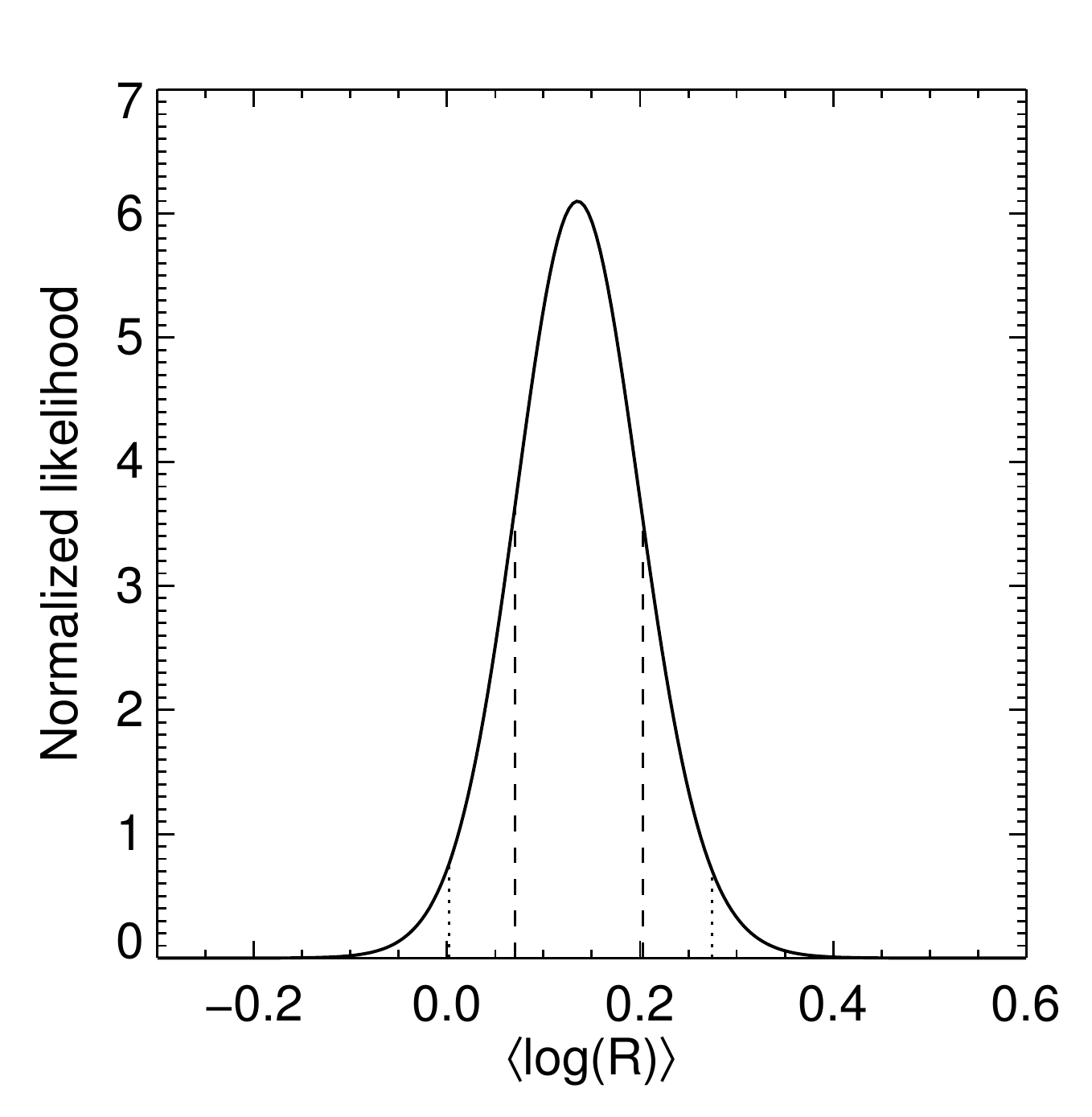}
	\caption{The normalized likelihood for the model presented in \S\ref{ExcitationAnalysis} as a function of the average flux ratio $X_0 = \langle \log(R) \rangle$, marginalized over $\sigma_0$.  Ranges corresponding to $1 \sigma$ and $2 \sigma$ are shown as dashed and dotted vertical lines, respectively.}
	\label{fig:likelihood_MaozMcKee}
\end{figure}

Once the joint likelihood function is known, we can marginalize over the parameter $\sigma_0$.  The marginalized likelihood, $\mathcal{L}(X_0)$ (shown in Figure \ref{fig:likelihood_MaozMcKee}), can then be integrated to determine the fraction of the likelihood that falls below $X_0 = 0$:

\begin{equation}
p = \left( \int_{-\infty}^{0} \mathcal{L}(X_0) dX_0 \right) \left( \int_{-\infty}^{\infty} \mathcal{L}(X_0) dX_0 \right)^{-1} . \label{eqn:Likelihood6}
\end{equation}

\noindent Evaluating $p$ for the flux values listed in Table \ref{tab:Maoz_McKee_targets} yields $p = 0.020$.  The likelihood analysis therefore rejects the null hypothesis at the 2$\sigma$ level.

NGC 4258 stands out as an 18$\sigma$ outlier, which most likely indicates that this Gaussian model is not a good description of the parent population.  Nevertheless, it is sufficient to show that the null hypothesis is at least moderately discrepant with the data and that NGC 4258 is substantially removed from the bulk of the observed distribution.

\subsection{Velocity drifts of high-velocity features} \label{ExcitationAccelerations}

The MM98 model also predicts that the high-velocity maser features will be systematically offset from the midline of the disk, and that they should thus exhibit nonzero line-of-sight velocity drifts as the spiral structure rotates.  For spiral shocks having a pitch angle of $\theta_p$ (the pitch angle is the opening angle of the spiral, defined at any point to be the complement of the angle between the tangent to the spiral and the outward radial direction from the black hole), we can calculate a characteristic value for the velocity drifts expected for the high-velocity features.  For a logarithmic spiral, the MM98 model predicts a velocity drift of

\begin{equation}
| \dot{v} | = 0.05 \left( \frac{\theta_p}{2.5^{\circ}} \right) \text{km s}^{-1} \text{ yr}^{-1} . \label{eqn:MM98Accelerations}
\end{equation}

\noindent This drift is towards smaller rotation velocities, and it is shared by all high-velocity masers.  The observed velocity drift, in this model, is caused by the passage of the trailing spiral structure through the gas disk; it is \textit{not} a centripetal acceleration from the Keplerian rotation of the gas.  As the spiral shock moves through the disk, the portion tangent to the line of sight intercepts gas farther out in radius, which has a lower rotational velocity.  Thus we would expect to observe a negative line-of-sight velocity drift for the redshifted features and a positive drift for the blueshifted features.

\cite{bra00} measured velocity drifts in NGC 4258, and showed that the values were inconsistent with the predictions of the MM98 model.  They established that no choice of pitch angle can reproduce their data, as statistically significant measurements of both negative and positive velocity drifts were made for both sets of features.  These results were corroborated by \cite{hum08}, who used an increased number of epochs to further refine the measurements.  Table \ref{tab:RB_accelerations} lists published measurements of high-velocity drifts for several other megamaser disks, plus our new measurements, where we have estimated velocity drifts for several additional galaxies using the eye-tracking method described in \cite{kuo13}.  To account for systematic uncertainties, we also adopt the error floor of 0.3 km s$^{-1}$ yr$^{-1}$ from \cite{kuo13} for all new acceleration measurements.

Nine of the 22 velocity drift measurements (counting redshifted and blueshifted separately) presented in Table \ref{tab:RB_accelerations} are incompatible with the MM98 model (i.e., negative blueshifted velocity drifts or positive redshifted velocity drifts).  For those values which are compatible, we used Equation \ref{eqn:MM98Accelerations} to assign a maximum pitch angle of any spiral structure that is consistent with the measured drifts.  As a comparison, the minimum pitch angle in NGC 4258 (obtained by assuming that the spatial grouping of the blueshifted features arise from consecutive windings of a single logarithmic spiral) is about $\theta_p \gtrsim 1.7^{\circ}$ \citep{hum13}.

\begin{deluxetable}{lcccc}
\tablecolumns{5}
\tablewidth{0pt}
\tablecaption{\label{tab:RB_accelerations}}
\tablehead{&	\colhead{Blue drifts} &	\colhead{Red drifts}	&	\colhead{$\theta_p$}	& \\
Target &	\colhead{(km s$^{-1}$ yr$^{-1}$)}	&	\colhead{(km s$^{-1}$ yr$^{-1}$)}	&	\colhead{(degrees)}	&	\colhead{Reference}}
\startdata
NGC 4258			&	$-0.140 \pm 0.03$\hphantom{0}								&	\hphantom{$-0$}$0.001 \pm 0.004$	&	$\ldots$			&	\cite{hum08}	\\
UGC 3789			&	$-0.046 \pm 0.04$\hphantom{0}								&	\hphantom{$-0$}$0.125 \pm 0.06$\hphantom{0}	&	$\ldots$			&	\cite{rei13}	\\
NGC 6264			&	\hphantom{$-$}$0.010 \pm 0.02$\hphantom{0}	&	\hphantom{0}$-0.130 \pm 0.01$\hphantom{0}		&	$< 0.50$ (b)		&	\cite{kuo13}	\\
NGC 6323			&	\hphantom{$-$}$0.030 \pm 0.15$\hphantom{0}	&	\hphantom{0}$-0.067 \pm 0.09$\hphantom{0}		&	$< 1.50$ (b)		&	\cite{kuo15}	\\
Mrk 1419			&	\hphantom{$-$}$0.007 \pm 0.14$\hphantom{0}	&	\hphantom{$-0$}$0.052 \pm 0.14$\hphantom{0}	&	$< 0.35$ (b)	&	$\ldots$	\\
NGC 1194			&	\hphantom{$-$}$0.031 \pm 0.13$\hphantom{0}	&	\hphantom{$-0$}$0.039 \pm 0.14$\hphantom{0}	&	$< 1.55$ (b)	&	Litzinger et al. (\textit{in prep.})	\\
NGC 2273			&	\hphantom{$-$}$0.074 \pm 0.23$\hphantom{0}	&	\hphantom{0}$-0.011 \pm 0.18$\hphantom{0}		&	$< 0.55$ (r)	&	$\ldots$	\\
J0437+2456		&	\hphantom{$-$}$0.036 \pm 0.14$\hphantom{0}	&	\hphantom{0}$-0.011 \pm 0.48$\hphantom{0}		&	$< 0.55$ (r)	&	$\ldots$	\\
ESO 558-G009	&	$-0.157 \pm 0.23$\hphantom{0}								&	\hphantom{0}$-0.047 \pm 0.22$\hphantom{0}		&	$< 6.25$ (r)	&	$\ldots$	\\
IC 2560				&	\hphantom{$-$}$0.011 \pm 0.15$\hphantom{0}	&	\hphantom{$0$}$-0.063 \pm 0.13$\hphantom{0}	&	$< 0.55$ (b)	&	$\ldots$	\\
NGC 5765b			&	$-0.049 \pm 0.04$\hphantom{0}								&	\hphantom{$-0$}$0.008 \pm 0.008$	&	$\ldots$			&	Gao et al. (\textit{submitted})	\\
\bottomrule \vspace{-3mm} \\
All						&	$-0.036 \pm 0.014$												&	\hphantom{$0$}$-0.012 \pm 0.003$	&	$< 0.60$ (b)			&		\\
\enddata
\tablecomments{This table lists the mean velocity drifts of high-velocity maser features in the best-sampled targets, along with their $1 \sigma$ statistical errors.  Values taken from the literature are accompanied by the appropriate citations; all other values are new measurements (see \S\ref{ExcitationAccelerations}).  Pitch angles are listed as upper limits, and they are calculated from the velocity drifts of either the redshifted (r) or blueshifted (b) features depending on which gives a tighter constraint.  Values incompatible with the MM98 model have no associated pitch angle.}
\end{deluxetable}

\subsection{Discussion} \label{ExcitationDiscussion}

Our analysis of the flux ratio data indicate a small deviation from the null hypothesis, in favor of the MM98 model.  However, the measured velocity drifts of high-velocity features do not match the MM98 predictions (Table \ref{tab:RB_accelerations}).  The maser features are equally likely to have a positive drift as a negative one, regardless of whether they're blueshifted or redshifted (6 of 11 targets display negative velocity drifts for both sets of features).  Furthermore, though we have reported only the averaged values for the redshifted and blueshifted velocity drifts for each target, several of these targets have statistically significant measurements of both negative and positive drifts within the same set of features.  On the whole, the high-velocity drifts are consistent with masing gas that is near the midline of the disk (i.e., any observed velocity drifts can be explained as centripetal accelerations caused by small offsets on both sides of the midline).

We note that the MM98 model is based on the characteristics of NGC 4258, which has an atypically large flux ratio between the redshifted and blueshifted high-velocity features.  This apparent anomaly could be the result of a selection bias.  If NGC 4258 were located at a distance of $\sim 100$ Mpc, which is more typical of our sample, it would likely not have been identified as a disk maser.  The systemic features would peak at about 25 mJy, and the strongest high-velocity features would only be about 3 mJy (i.e., marginally detectable in a single-epoch GBT spectrum).  However, it is also true that our selection criteria (see \S\ref{DiskSpectra}) allowed for the presence of highly asymmetric flux ratios in the sample (e.g., an NGC 4258 analogue at a distance of 50 Mpc), yet we found none other than NGC 4258 itself.  As such, we retain the assertion that NGC 4258 is truly anomalous in having such a large flux ratio.

\section{Variability} \label{Variability}

There are several classes of variability present in the megamaser spectra, with different timescales and presumed underlying physical causes.  We qualitatively outline these classes in this section.

Long-term ($\sim$hundreds of days) ``bulk variability" in the line flux of maser feature sets is seen in all sufficiently monitored galaxies.  The dynamical timescale for a $\sim$1 pc accretion disk around a $\sim$$10^7$ M$_{\odot}$ black hole is $\sim$$10^4$ years, so if this bulk variability has a dynamical origin, then it likely originates from activity much closer to the central AGN than any observed masers.  \cite{gal01} argue that the megamasers in NGC 1068 respond to changes in the central power source, via a reverberation mechanism.  We investigate this possibility for several other galaxies in \S\ref{DiskReverberation}.

Many maser galaxies also display short-term ($\sim$monthly) flaring variability, where a single maser line increases enormously in amplitude, often by several orders of magnitude over the course of only $\sim$a week and lasts for a few weeks.  This flaring may be caused by the chance alignments of individual masing gas clumps in the disk (see, e.g., \citealt{kar99}).  In this picture, masing occurs in localized clouds which are orbiting ballistically in the accretion disk.  When one cloud passes in front of another while maintaining velocity coherence (as might happen, e.g., for two high-velocity clouds on either side of the disk midline), the foreground cloud further amplifies the emission from the background cloud, resulting in a rapid increase in line luminosity.  This provides another potential mechanism for the bulk variability, as it could be the combined flares of many weak, blended maser lines.

Extremely short-term (intra-day) variability that is also uncorrelated among different spectral features has been observed in two megamaser galaxies: Circinus \citep{mcc05} and NGC 3079 \citep{vle07}.  This variability has been attributed to interstellar scintillation, and in \S\ref{Scintillation} we present evidence for such scintillation in a third megamaser galaxy, ESO 558-G009.

We note that our observations are only sensitive to variability on $\lesssim$hourly timescales and $\gtrsim$monthly timescales.

\subsection{Dynamic spectra} \label{DynamicSpectra}

\begin{figure*}[p]
	\centering
		\subfigure[UGC 3789]{\includegraphics[width=1.00\textwidth]{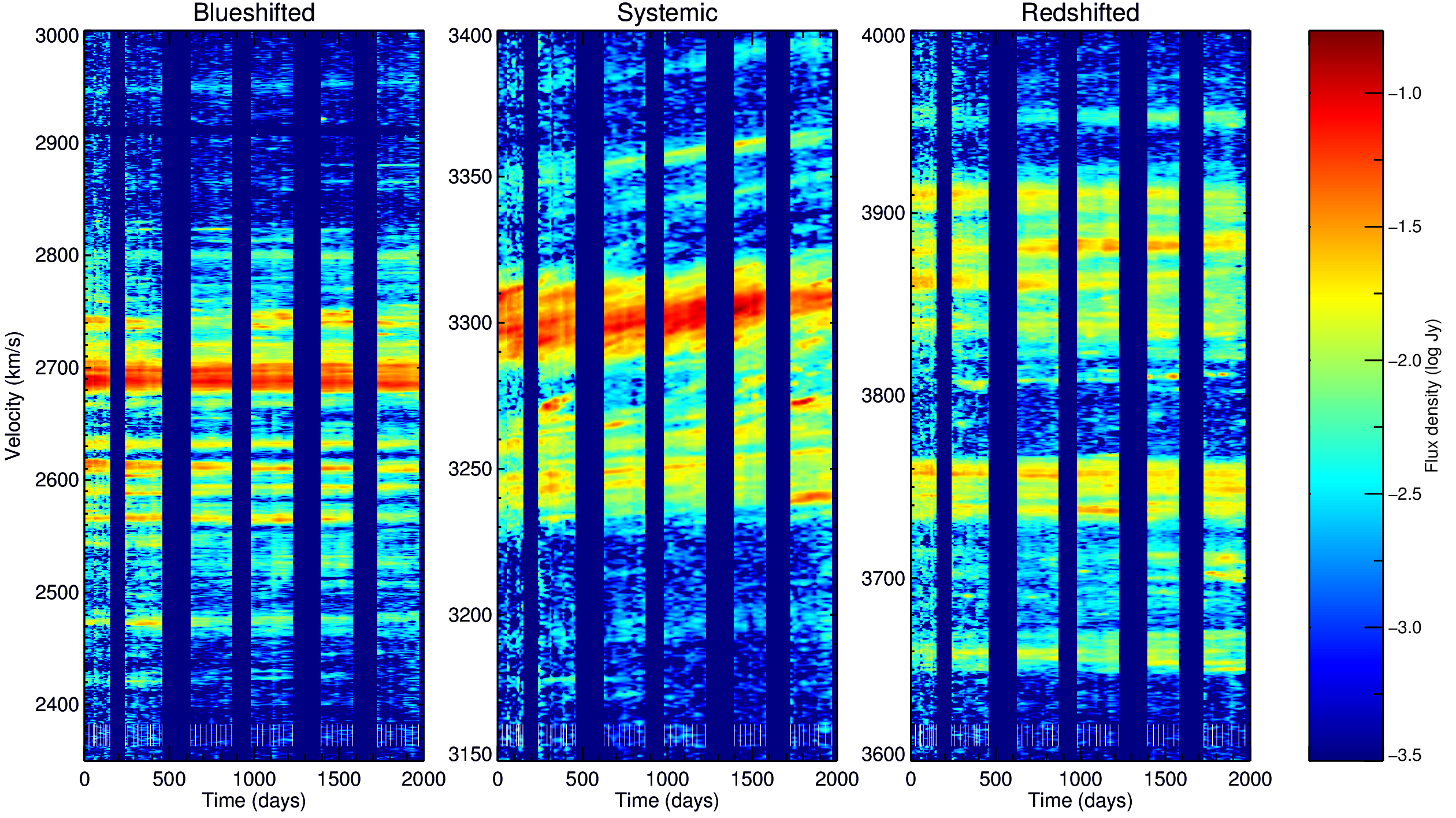} \label{fig:UGC3789_dynam_spec_subfig}}
		\caption{Dynamic spectra for our best-sampled disk megamasers.  For ease of viewing, the three sets of features have been split up and the spectral regions in between (which are devoid of maser features) are not shown.  The color scale maps to the logarithm of the flux density, as shown in the colorbar on the right.  Individual observation dates are indicated by white tick marks near the bottom of each plot, and day zero is set as the date of the first observation (see Figure \ref{fig:AveragedSpectra}).  Velocities are measured in the heliocentric frame, using the optical velocity convention.} \label{fig:dynamic_spectra}
\end{figure*}

One way to effectively visualize both the bulk variability and the flaring variability is through dynamic spectra.  In Figure \ref{fig:dynamic_spectra} we present dynamic spectra for 9 of our best-sampled sources.  To create the dynamic spectra, we linearly interpolated the flux densities between consecutive GBT spectra, which were taken at a roughly monthly cadence.  For the MCP's monitoring campaign, targets were not observed during the North American summer because atmospheric conditions in Green Bank make K-band observations inefficient during this season.  Summer periods with no data are blanked.

Kinematic differences between the systemic and high-velocity features, corresponding to differences in line-of-sight accelerations, are immediately apparent in the dynamic spectra.  Figures \ref{fig:dynamic_spectra}\subref{fig:UGC3789_dynam_spec_subfig} and \ref{fig:dynamic_spectra}\subref{fig:NGC6264_dynam_spec_subfig} match well with Fig. 2 from \cite{bra10} and Fig. 1 from \cite{kuo13}, respectively.  Further, we note that the systemic feature located initially at $\sim$3380 km s$^{-1}$ in UGC 3789, which was not used for the distance determination by \cite{rei13} in their acceleration analysis for signal-to-noise reasons, shows a clear acceleration in the dynamic spectrum.  This feature is offset by about 15 km s$^{-1}$ from the nearest systemic features for which an acceleration was measured, so including it would expand the velocity span of the systemic feature set by $\sim$12\% and potentially improve the disk model and associated distance measurement.

Along with the kinematic information, the dynamic spectra also illustrate how the flux densities and overall spectral shape change with time.  If we follow, for instance, the systemic features at $\sim$3270 km s$^{-1}$ in UGC 3789, we can see that they vary in amplitude by more than an order of magnitude during the $\sim$6-year span of these observations.  We can also see features near this velocity appearing and disappearing with time.  Several of the blueshifted features bracketing 2600 km s$^{-1}$, on the other hand, remain quite stable in both amplitude and structure during the same time range.  There are also marked differences in feature stability among different galaxies; NGC 5765b, for instance, has a very consistent spectrum compared to the others.  As a result of this spectral stability, NGC 5765b has the most precisely-measured distance of any MCP galaxy to date (Gao et al. \textit{submitted}).  NGC 1194, on the other hand, is observed to be extremely variable; this variability has made measurements of this galaxy very challenging (Litzinger et al. \textit{in prep}).

Additionally, we can compare the lifetimes of different flaring features in the spectra.  The 3270 km s$^{-1}$ systemic feature in UGC 3789 flared at around day 1700, and it lasted roughly 200 days.  This duration is considerably longer than that of the 3810 km s$^{-1}$ redshifted feature, which flared around day 1500 but only lasted $\sim$50 days.  Compare this to the 1580 km s$^{-1}$ feature in NGC 2273, which lasted for at least 400 days, and the 8005 km s$^{-1}$ feature in ESO 558-G009, which had a duration of $\sim$100 days.

\begin{figure*}[p]
	\centering
		\addtocounter{subfigure}{1}
		\captcont{(continued)}
		\subfigure[ESO 558-G009]{\includegraphics[width=1.00\textwidth]{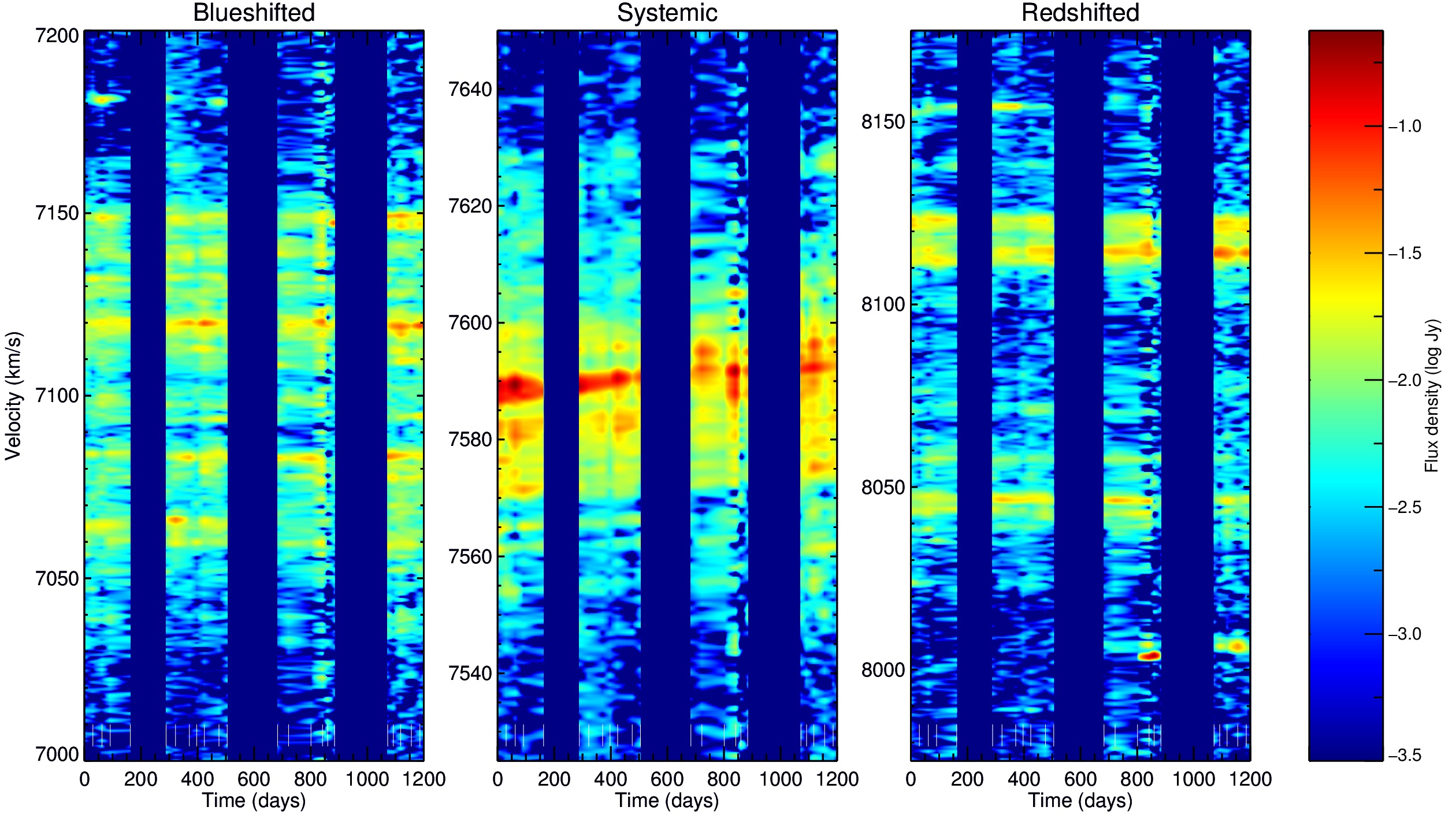} \label{fig:ESO558-G009_dynam_spec_subfig}} \\
		\subfigure[J0437+2456]{\includegraphics[width=1.00\textwidth]{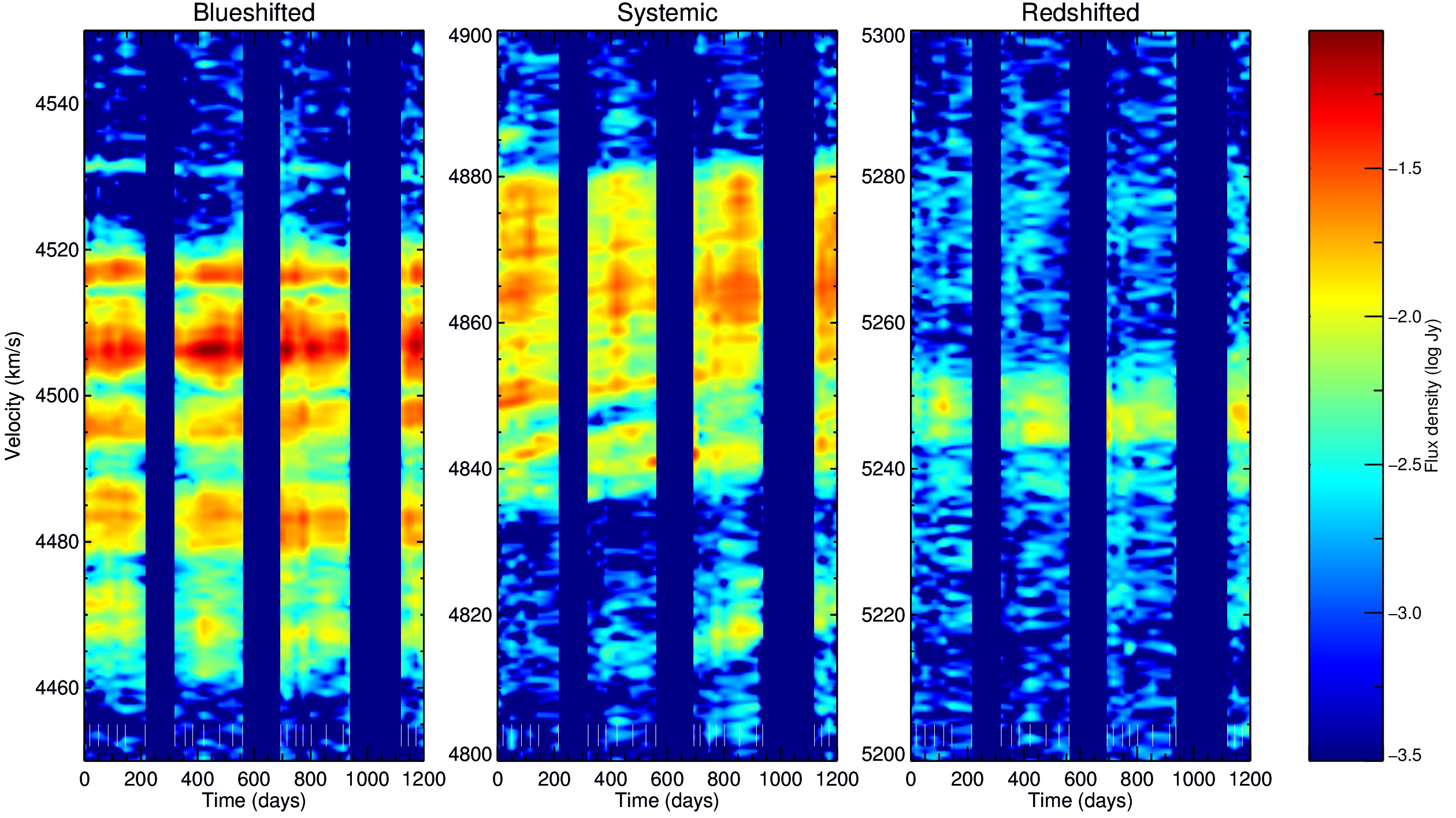} \label{fig:J0437+2456_dynam_spec_subfig}}
\end{figure*}
\begin{figure*}[p]
	\centering
		\captcont{(continued)}
		\subfigure[Mrk 1419]{\includegraphics[width=1.00\textwidth]{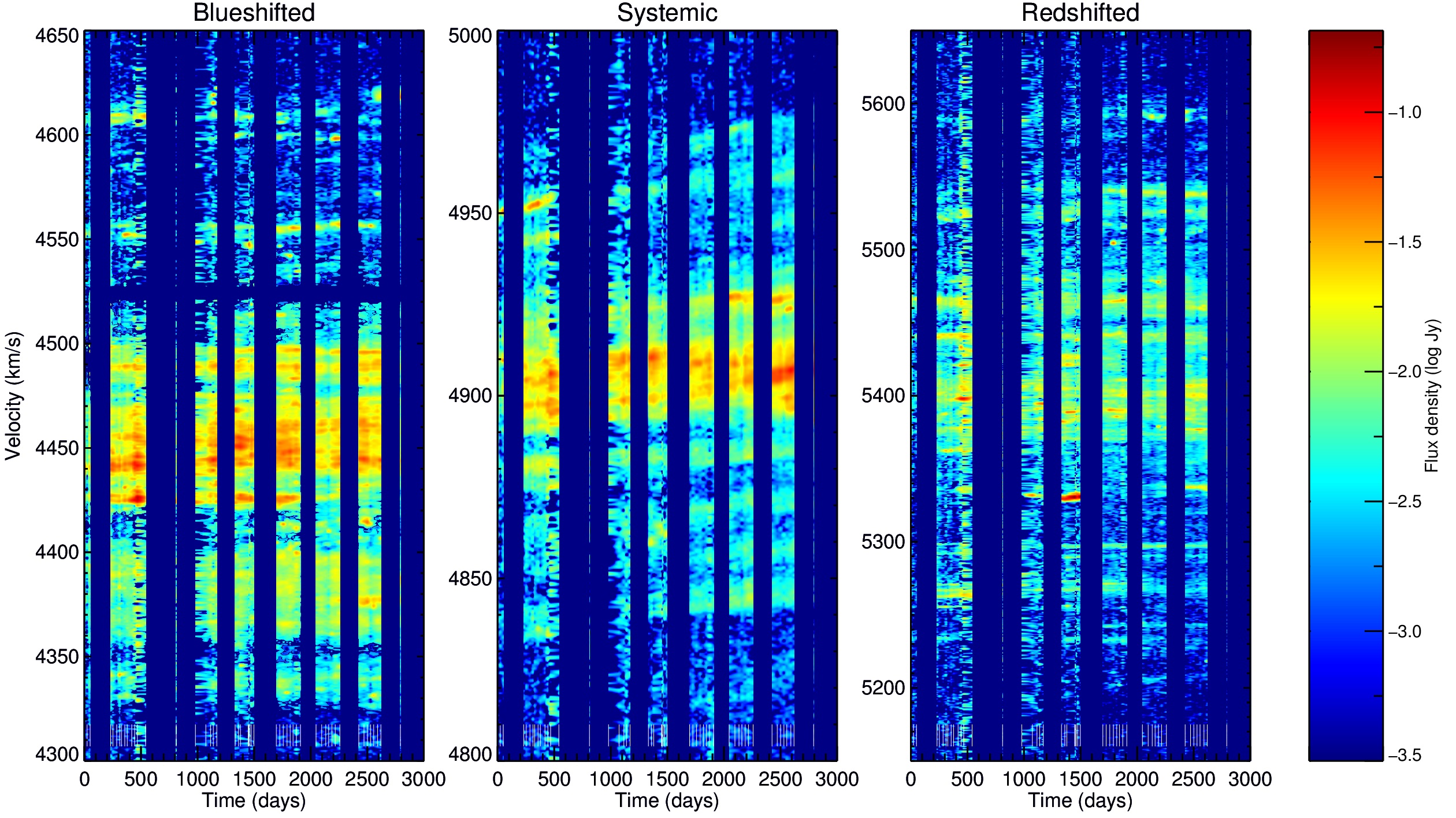} \label{fig:Mrk1419_dynam_spec_subfig}} \\
		\subfigure[NGC 1194]{\includegraphics[width=1.00\textwidth]{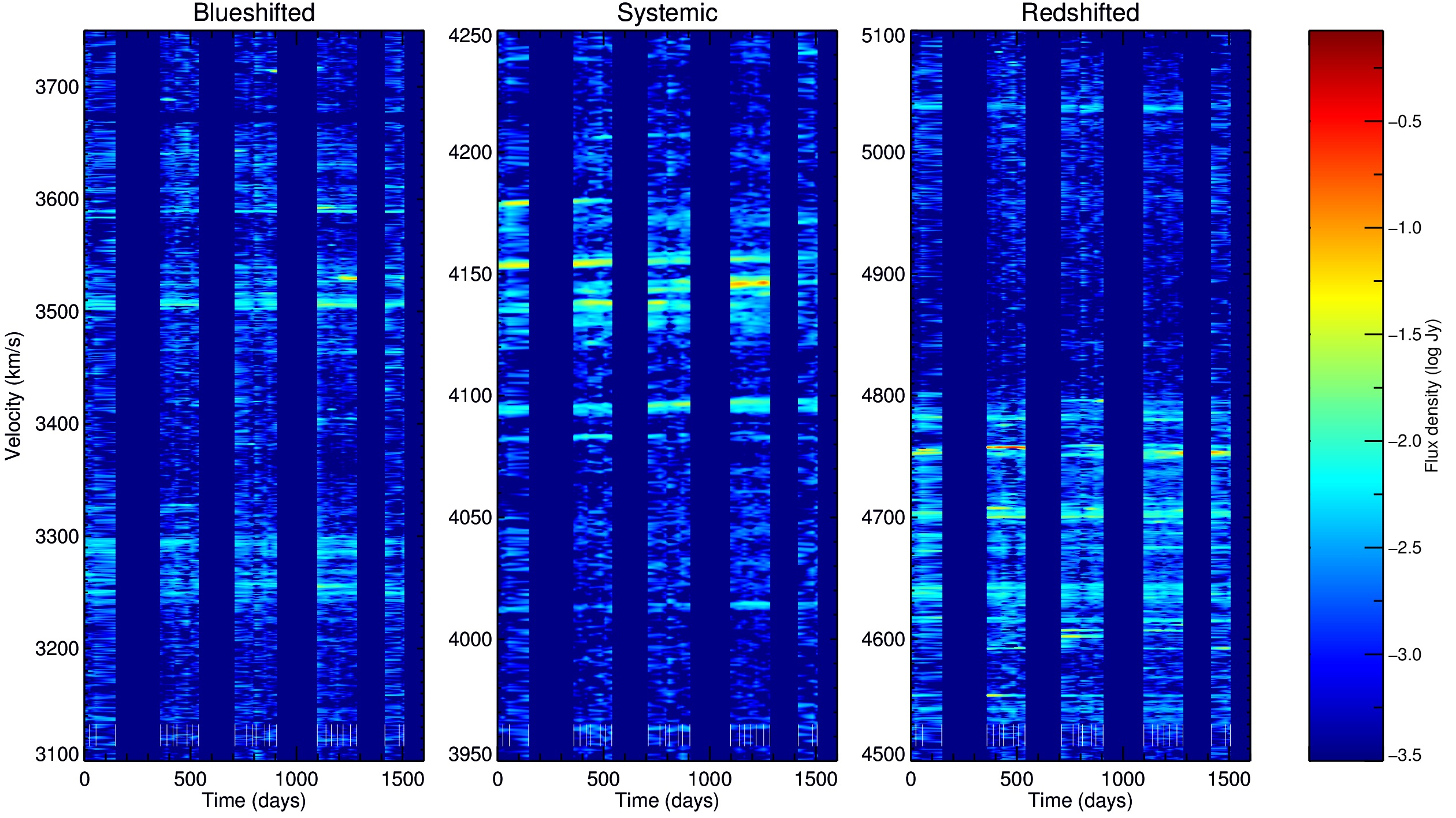} \label{fig:NGC1194_dynam_spec_subfig}}
\end{figure*}
\begin{figure*}[p]
	\centering
		\captcont{(continued)}
		\subfigure[NGC 2273]{\includegraphics[width=1.00\textwidth]{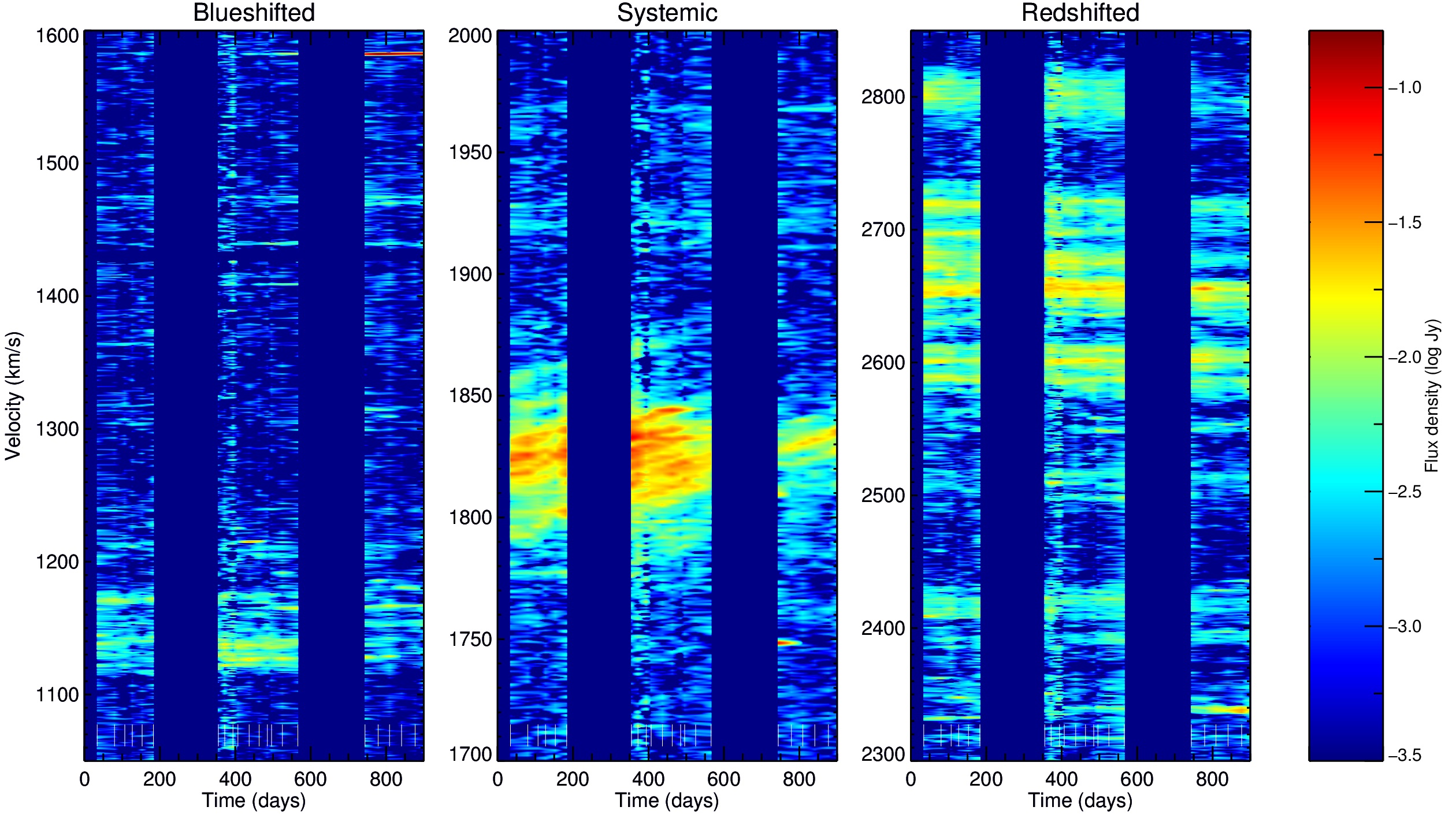} \label{fig:NGC2273_dynam_spec_subfig}} \\
		\subfigure[NGC 5765b]{\includegraphics[width=1.00\textwidth]{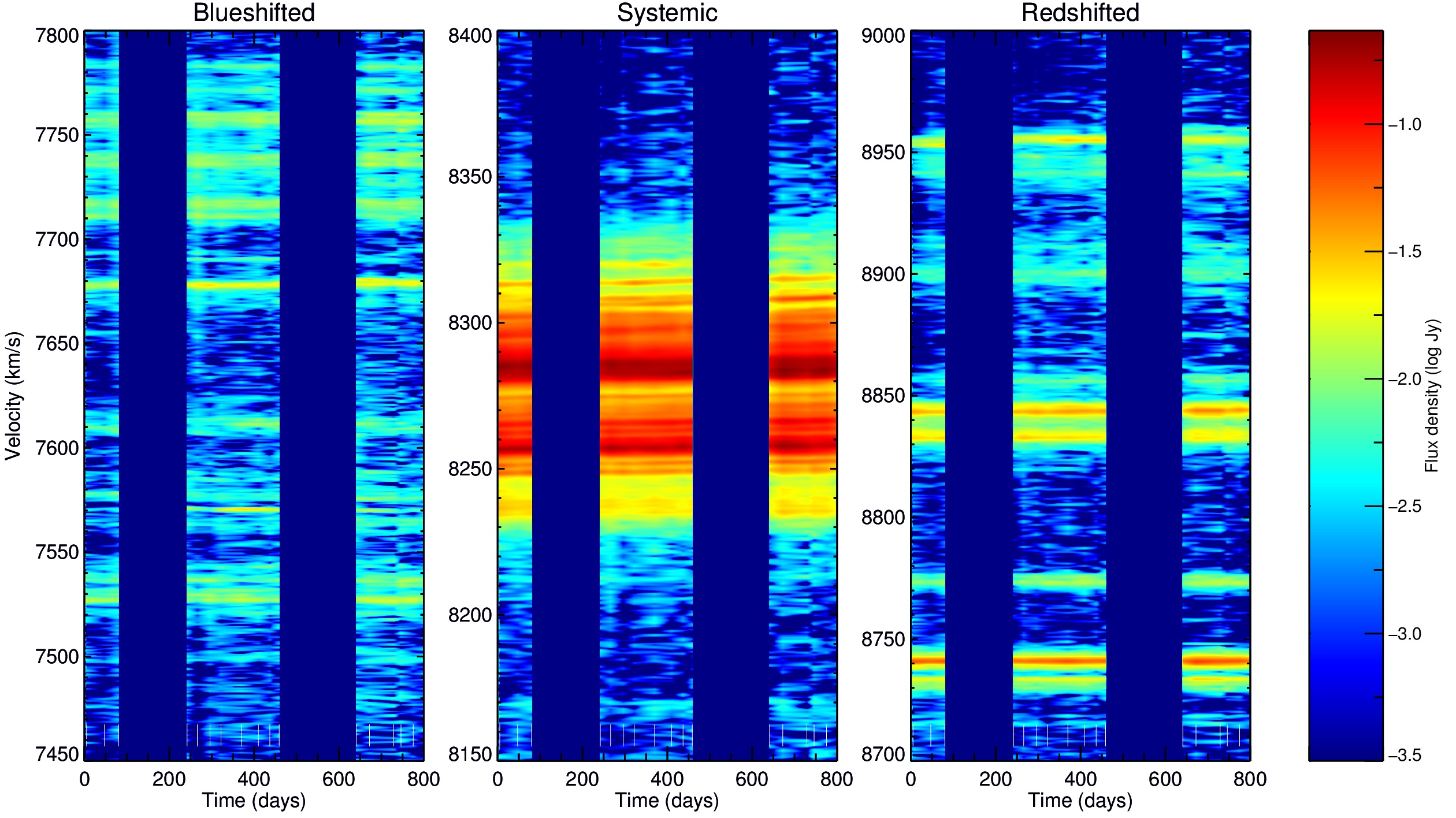} \label{fig:NGC5765b_dynam_spec_subfig}}
\end{figure*}
\begin{figure*}[p]
	\centering
		\captcont{(continued)}
		\subfigure[NGC 6264]{\includegraphics[width=1.00\textwidth]{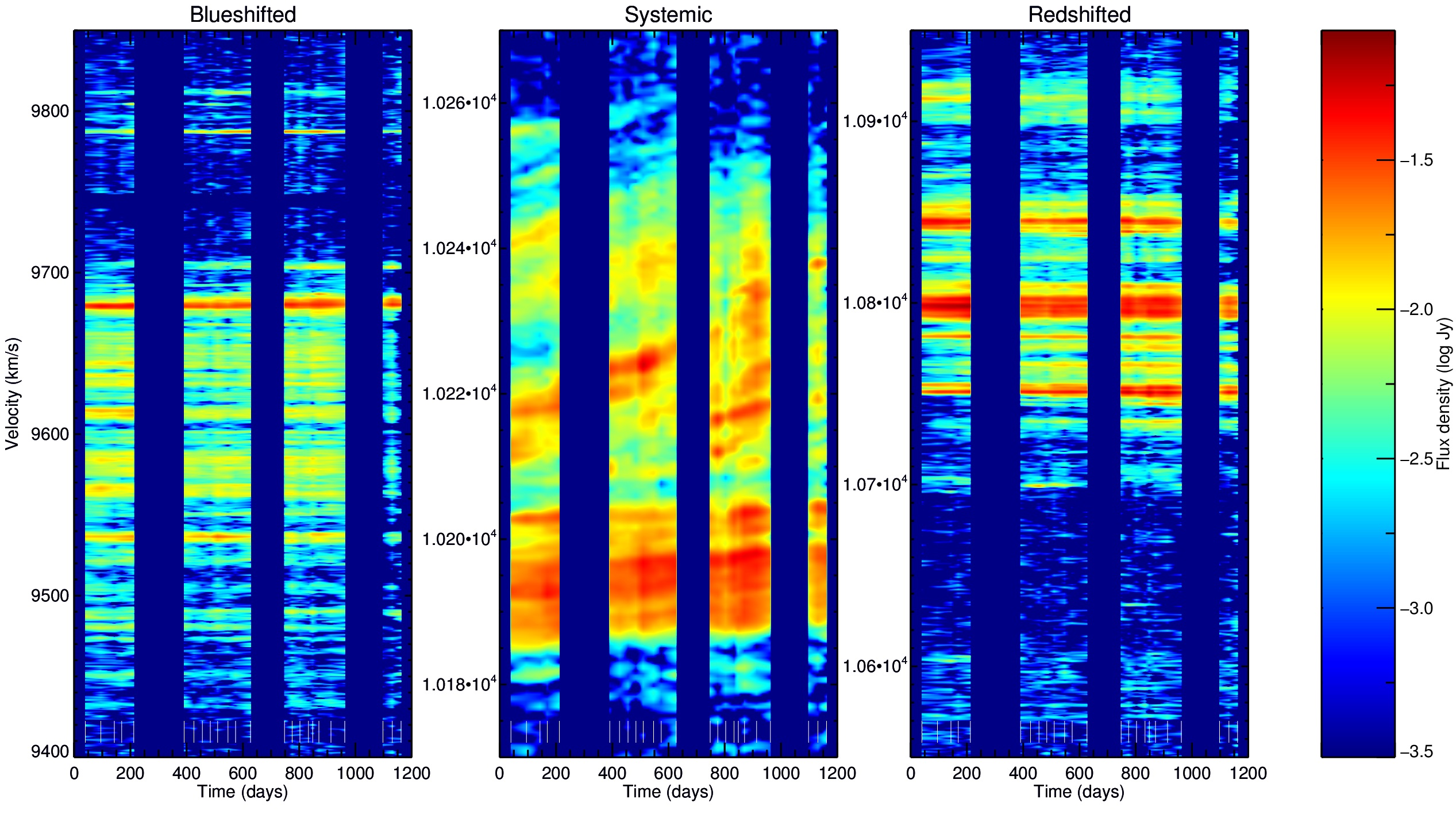} \label{fig:NGC6264_dynam_spec_subfig}} \\
		\subfigure[NGC 6323]{\includegraphics[width=1.00\textwidth]{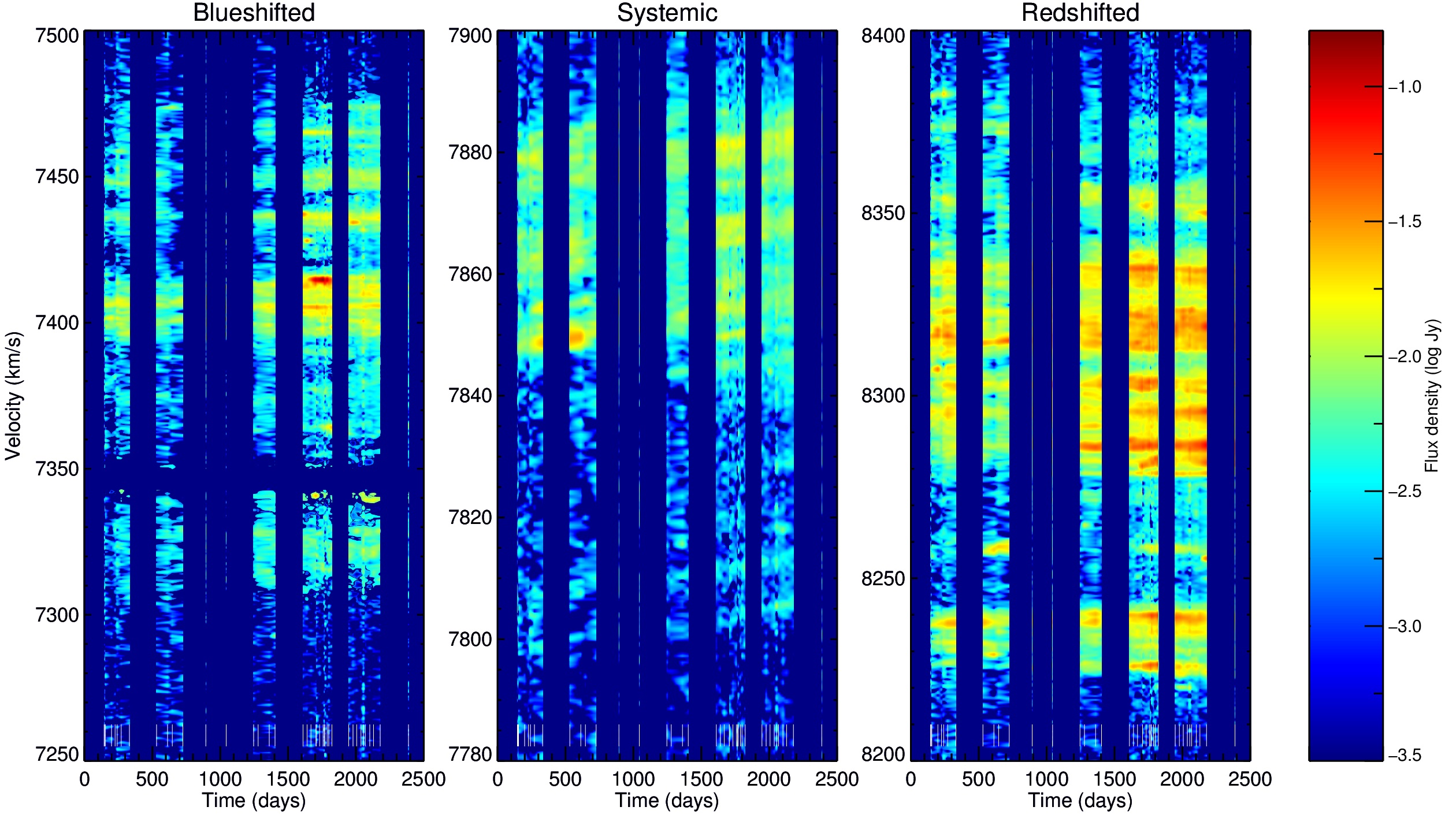} \label{fig:NGC6323_dynam_spec_subfig}}
\end{figure*}

\subsection{Scintillation} \label{Scintillation}

Interstellar scintillation (ISS) in the Galactic ionized ISM is considered to be the primary mechanism causing the rapid intraday variability observed in pulsars and many extragalactic radio sources (predominantly quasars; see, e.g., \citealt{big04}).  For a distant source whose emission is undergoing scattering in the turbulent ISM of our Galaxy, it is simplest to treat the sum contribution from the line-of-sight electron column as originating from a single thin ``scattering screen" located a distance $D$ from the Earth.  In this picture, turbulence is generated on timescales that are much longer than the time it takes a phase-coherent region of the scattering medium (dubbed a ``scintle") to cross the source.  That is, the phase variations introduced by the screen are essentially ``frozen" as the screen passes across the line of sight.  Thus, the scintillation timescale is set by the size and transverse velocity of the scattering scintle.

There are two important ISS regimes separated by a ``transition frequency" $\nu_t$: the weak ($\nu > \nu_t$) and strong ($\nu < \nu_t$) scattering limits.  We give a brief overview of some relevant properties of these limits here; for a review of this topic, see \cite{nar92} and references therein.

In the weak scattering limit, the size of the scintle is of order the Fresnel scale, defined to be the transverse distance from the line of sight to a point through which the increase in path length from the source to observer (compared to the direct, line-of-sight path) results in a phase change of 1 radian.  For a source at infinity and an observing wavelength $\lambda \ll D$, the Fresnel scale is given by $r_{\text{F}} = \sqrt{\lambda D / 2 \pi}$.  If the scattering screen has transverse velocity (relative to the Earth) of $v$, the variability timescale will be $\tau \approx r_{\text{F}} / v$.

In the strong scattering limit, the scintle has a characteristic size called the diffractive scale, $r_{\text{diff}}$.  This length scale functions equivalently to the Fresnel scale in weak scintillation (i.e., the RMS phase difference between two points on the screen separated by a distance $r_{\text{diff}}$ is approximately 1 radian), but the physical origin of the size scale is different.  In the strong scattering regime, the value of $r_{\text{diff}}$ is determined by the turbulent properties of the ISM plasma rather than by the geometry of the observer-screen-source setup.  We thus have $r_{\text{diff}} \ll r_{\text{F}}$ for strong scattering, while $r_{\text{F}} \ll r_{\text{diff}}$ for weak scattering.  The scintillation timescale will then be $\tau \approx r_{\text{diff}}/v$.  The strong scattering regime can be further subdivided into two different types of strong scattering, diffractive and refractive.  Refractive scintillation occurs on much longer timescales ($\sim$days) than diffractive scintillation, so it is not relevant for this study.

A standard measure of variability strength is the modulation index, $\mu = \sigma / \langle S \rangle$, where $\sigma$ is the standard deviation of the observed amplitude and $\langle S \rangle$ is its average value.  The modulation index for a point source undergoing weak scattering is roughly the ratio of the Fresnel to the diffractive scale, $\mu \approx (r_{\text{F}} / r_{\text{diff}})^{5/6}$ \citep{nar92}.  For diffractive scintillation, the modulation index should be unity.  In the case of an extended source (i.e., a source with an angular size larger than the diffractive scale), the diffractive scintillation is said to be ``quenched," since the resolved source is effectively diluting the variability amplitude by averaging the phase fluctuations over several adjacent scintles.  An extended source of angular size $\theta$ will have a modulation index given by $\theta_{\text{diff}} / \theta$.

ISS has been proposed as an explanation for the extremely rapid (intra-hour) variability observed in the 22 GHz maser spectra from the Circinus galaxy and NGC 3079.  \cite{vle07} use the high Galactic latitude of NGC 3079 ($b = +$48.36$^{\circ}$) to justify their assumption of weak scintillation.  From a measured characteristic timescale of $\tau \approx 1000$ s, corresponding to the crossing time for the Fresnel scale, they calculate a distance to the scattering screen of $D \approx 25$ pc.

\cite{mcc05} measured the timescale in Circinus to be $\tau \approx 700$ s, but were unable to say definitively whether the variability was caused by weak scintillation in a nearby screen ($D \approx 20$ pc) or quenched diffractive scintillation in a more distant screen ($D \approx 230$-1000 pc).  Followup observations from \cite{mcc07} showed spectral variations that lent strong support to the diffractive scintillation interpretation, and they further uncovered longer-timescale ($\sim$1 day) variations consistent with refractive scintillation.

\subsubsection{Scintillation in ESO 558-G009} \label{ScintillationESO558}

\begin{figure*}
		\includegraphics[width=1.00\textwidth]{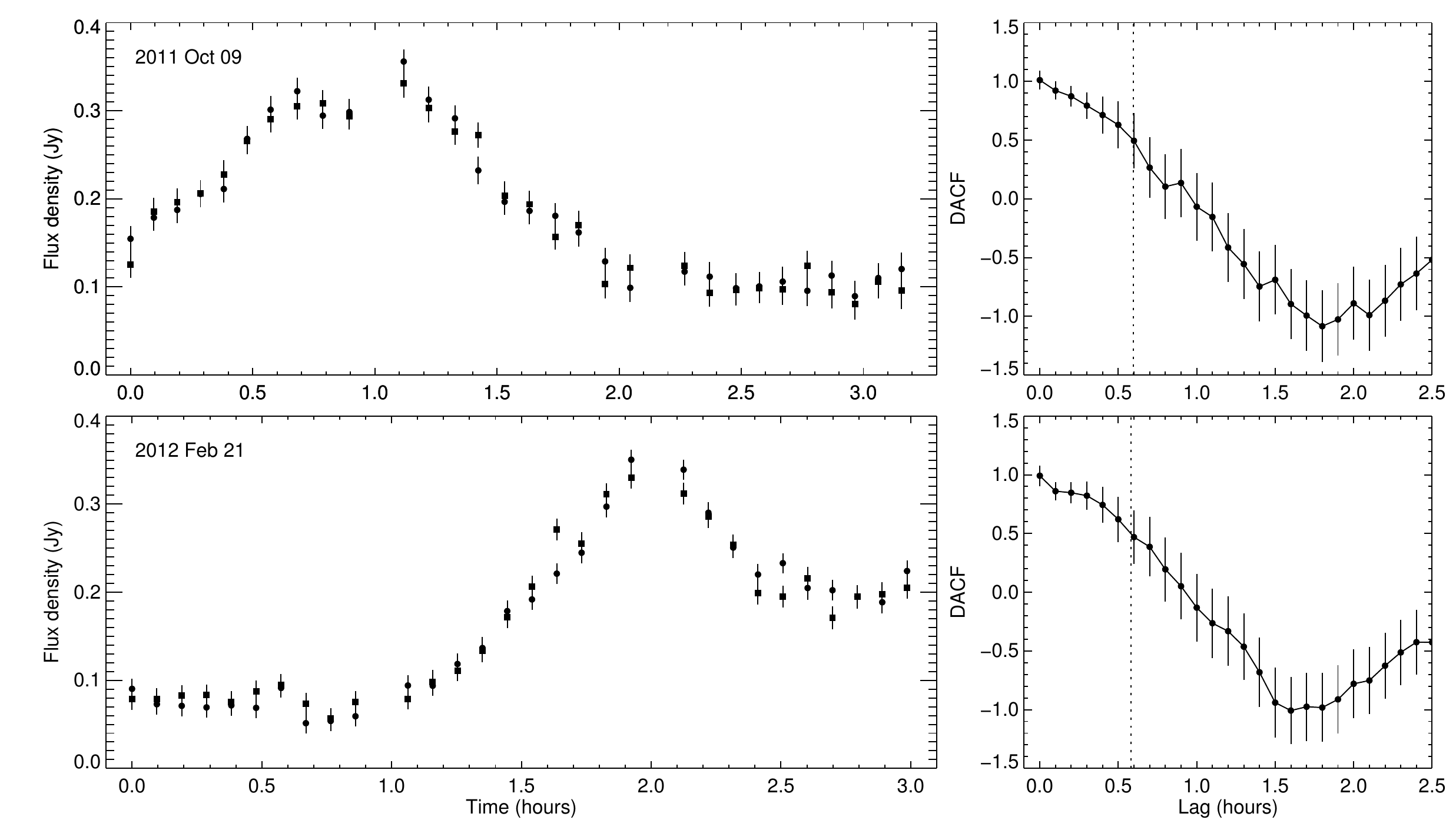}
	\caption{Light curves (left) and discrete autocorrelation functions (DACF, right) for two observations of ESO 558-G009.  In the light curves, the LCP and RCP peak flux densities of the 7590 km s$^{-1}$ maser line are plotted (with circles and squares, respectively) at the $\sim 5$-minute cadence corresponding to individual nod scan pairs.  The dotted vertical line in the DACF marks the location of $\tau$ (i.e., where the DACF drops to a value of 0.5).}
	\label{fig:scintillation}
\end{figure*}

We present here observations of the third megamaser galaxy observed to show signs of ISS.  Figure \ref{fig:scintillation} shows light curves for two epochs of the galaxy ESO 558-G009 on which we've applied our scintillation analysis.  These epochs were chosen because of their long observation durations ($\gtrsim 3$ hours each) and because they both contained the same strong systemic maser feature ($\gtrsim 0.15$ Jy), which was detectable in a single 5-minute scan.  We examined the spectra for all the other megamasers that met these same criteria (long-duration observation and strong maser feature), but only ESO 558-G009 showed significant variability.  Figure \ref{fig:scintillation} also shows the discrete autocorrelation functions (DACFs) for both of the light curves, calculated using the technique outlined by \cite{ede88}.  The dates of the observations are listed in Table \ref{tab:scintillation}.

The light curves show variability timescales on the order of $\sim$2100 s, during which the peak flux can vary by a factor of $\sim$3;  this is comparable to the amplitude modulations observed in the quasar J1819+3845, the extragalactic source exhibiting the strongest ISS-induced continuum variability \citep{den02}.  Though it's possible for pointing errors or a changing atmospheric opacity to cause the amplitudes of spectral features to vary with time, we don't expect these effects to exceed $\sim$20\%.  Further, if such factors were the cause of the observed amplitude changes then we would expect to see them across spectra of all galaxies, which is not the case.  As a final check, we measured the total flux of the systemic and high-velocity features (outside of the targeted line) over time during the observations, and we found that it is constant to within $\sim$15\% throughout a single observing session.

If the variability were intrinsic to the maser, then such large amplitude changes must result from increases in the maser gain path that are of order the gain length, $\ell$, which for an unsaturated maser is the path length corresponding to an $e$-fold increase in amplification (i.e., it is the length over which the optical depth $\tau$ changes by $\sim$1).  For conditions typical of those found in megamaser disks, $\ell \gg 1$ AU \citep{gre97a}.  Given the light-travel distance of $\sim$4 AU derived from the characteristic timescale, the observed variability would require changes in the gain path to propagate at approximately the speed of light.  Barring radiative pumping (which is not expected to be important in these systems; see, e.g., \citealt{lo05}), we do not know of any mechanism capable of driving such rapid changes.  This leaves foreground scintillation as the best available explanation for the variability.

Following \cite{ric02}, we define the characteristic observed scintillation timescale, $\tau$, to be the half-width at half-maximum (HWHM) of the autocorrelation function.  If the masers behave as a point source (i.e., if their angular size is smaller than the angular size of the scattering screen, $\theta < \theta_s$), then a measurement of $\tau$ allows us to establish a characteristic size, $r_s$, for the scattering screen (i.e., the size of a scintle) of

\begin{equation}
r_s = v_s \tau .
\end{equation}

\noindent Here, $v_s$ is the transverse velocity of the screen relative to Earth.  In Appendix \ref{app:TransverseMotion} we have outlined how this transverse velocity is obtained for an individual observation, using a model that combines the Earth's orbital motion and the Sun's peculiar and orbital motion.  Table \ref{tab:scintillation} lists $v_s$ for each observation, assuming a nearby ($D \lesssim 100$ pc) screen; the measured values for $\tau$ are also listed.

\begin{figure*}
		\includegraphics[width=1.00\textwidth]{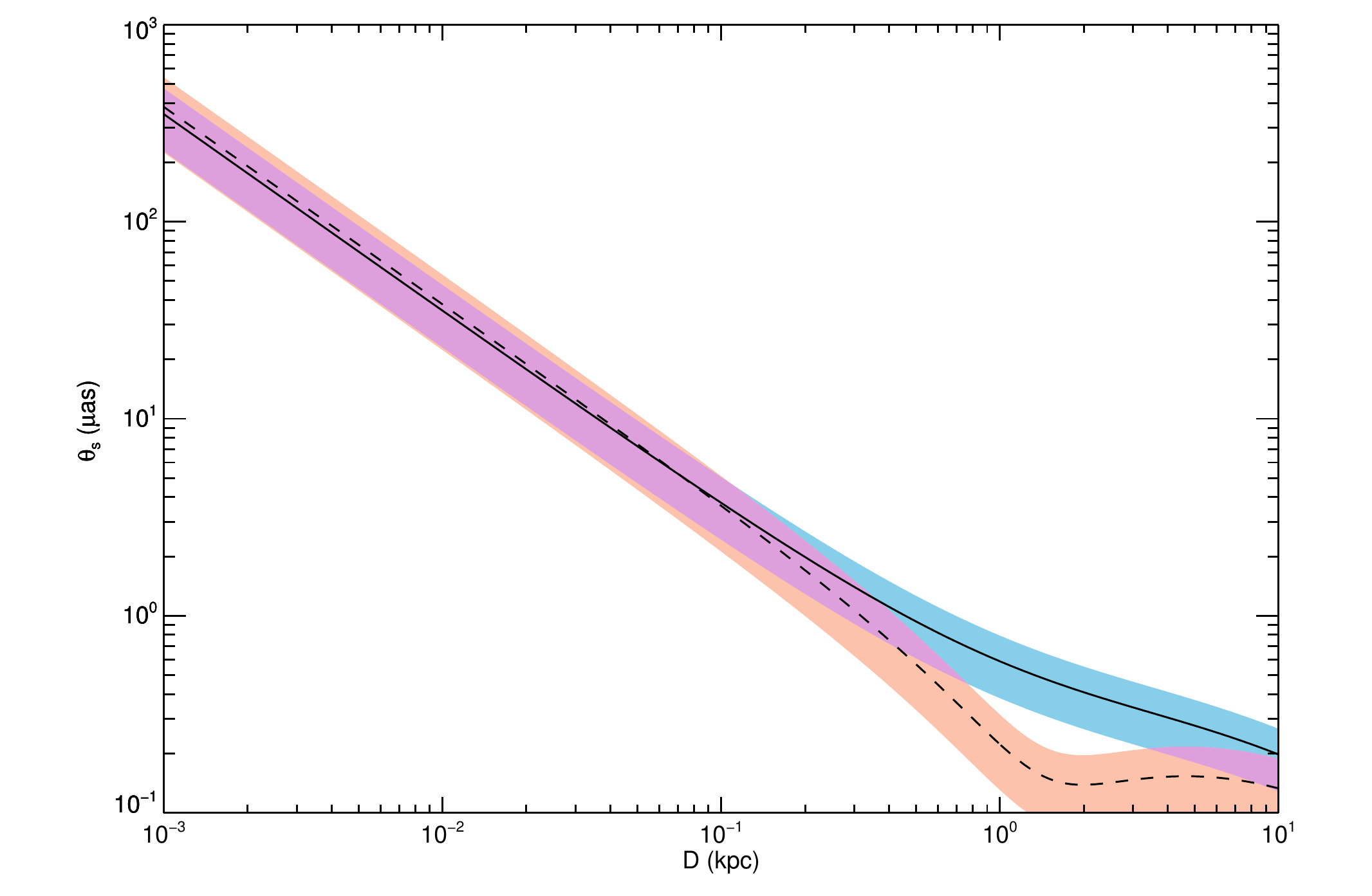}
	\caption{Constraints on the angular size of the scintles and the distance to the scattering screen along the line of sight to ESO 558-G009.  The solid line (with 3$\sigma$ error in blue) shows the constraint from the 2011 October 9 observation, and the dashed line (with 3$\sigma$ error in red) shows the constraint from the 2012 February 21 observation.  The plotted errors account only for the statistical errors arising from the measurement of $\tau$; no systematic errors from the velocity modeling are included.}
	\label{fig:scintle_fit}
\end{figure*}

Our model assumes that the scattering screen itself has no peculiar motion.  From time-delay measurements of the intra-day variability in the quasar J1819+3845, \cite{den02} found that the scattering screen (for that target) must have a transverse peculiar velocity of about 25 km s$^{-1}$.  \cite{mcc09} used the same technique to place a lower limit of 22 km s$^{-1}$ on the transverse velocity of the ISM along the line of sight to Circinus.  We have no reason to expect that the scattering screen towards ESO 558-G009 should behave any differently.  However, with two free parameters already in the model ($D$ and $r_s$) and only two measurements, we have no room to add the two additional parameters that would be necessary to properly account for peculiar motion.  We are thus only able to place relatively crude constraints on the model parameters.  Figure \ref{fig:scintle_fit} shows these constraints, with the more relevant $\theta_s = r_s / D$ plotted in place of $r_s$.  We can see that our measurements, which have a formal ``best fit" at about $D \approx 70$ pc and $\theta_s \approx 5$ $\mu$as, are compatible with a wide range of parameters.  A scintle angular size of 5 $\mu$as corresponds to a lower limit on the maser brightness temperature of $\sim$$3 \times 10^{13}$ K.

If we assume that the scintillation occurs in the weak scattering regime, then we have $r_s \approx r_{\text{F}}$, and we can use the Fresnel scale to determine $D$.  Doing so yields a distance to the scattering screen between 40 and 50 pc.  From \cite{wal98}, we can use the modulation index to determine the transition frequency.  Between the two observations, $\mu \approx 0.5$, so we obtain $\nu_t \approx 13.6$ GHz.

Like Circinus, ESO 558-G009 is located near the plane of the Galaxy ($b = -6.96^{\circ}$), so we would expect to see greater-than-average scattering along this line of sight.  From the NE2001 model for the electron density along different lines of sight in the Milky Way \citep{cor02}, the transition frequency between weak and strong scintillation towards ESO 558-G009 should actually be about 30 GHz; since this is higher than the observing frequency of 22 GHz, it would put us in the strong limit.  We note that the \cite{cor02} model attempts to map the Galactic electron density in a primarily spatially smooth manner, while the true distribution is known to have mesoscale and microscale structure.  As such, we expect significant model uncertainties along any specific line of sight.

In the strong scintillation regime the measured timescale maps to the angular size of the source rather than to that of a scintle.  The modulation index should be equal to the ratio $\theta_{\text{diff}} / \theta_{\text{s}}$, so a modulation index of $\mu \approx 0.5$ (see Table \ref{tab:scintillation}) indicates that the angular size of the maser must be a factor of $\sim$2 larger than that of the scintle.  For a screen distance of 70 pc we have $\theta_{\text{s}} \approx 5$ $\mu$as.  For the ESO 558-G009 distance of 110 Mpc, we thus obtain an approximate physical size of the masing region of $\sim$1100 AU.  This is comparable to the 0.001--0.006 pc clump sizes estimated by \cite{kon05} for the disk of NGC 3079.

\begin{deluxetable}{lcccccc}
\tablecolumns{7}
\tablewidth{0pt}
\tablecaption{\label{tab:scintillation}}
\tablehead{	&	\colhead{$v$}	&	\colhead{$v_s$} &	\colhead{$\langle S \rangle$}	&	\colhead{$\sigma$}	&		&	\colhead{$\tau$} \\
Date	&	\colhead{(km s$^{-1}$)} &	\colhead{(km s$^{-1}$)}	&	\colhead{(Jy)}	&	\colhead{(Jy)}	&	\colhead{$\mu$}	&	\colhead{(hours)}}
\startdata
2011 Oct 09	&	7589.8		&	24.4	&	0.186	&	0.082	&	0.44	&	$0.60 \pm 0.07$	\\
2012 Feb 21	&	7590.3		&	27.4	&	0.166	&	0.088	&	0.53	&	$0.58 \pm 0.08$	\\
\enddata
\tablecomments{Scintillation parameters for ESO 558-G009.  The column titled $v$ lists the Doppler velocity for the targeted line, $v_s$ is the modeled transverse velocity at the observation date (see Appendix \ref{app:TransverseMotion} for details), $\langle S \rangle$ is the mean flux density for the line during the observation, $\sigma$ is its standard deviation, $\mu$ is the modulation index, and $\tau$ is the measured characteristic variability time.}
\end{deluxetable}

\section{Testing for disk reverberation} \label{DiskReverberation}

\cite{cla86} noted that the apparent systematic flux variations in the nuclear masers in NGC 1068 suggested that the masers share a common pumping source.  If all masers in a given galaxy are powered by a common source, presumably at the nucleus, then we would expect variability in the power source to propagate to the maser system.  This variation would reverberate through the masing disk at some propagation velocity which, if it is on the order of the speed of light, would be fast enough to pass through the entire masing portion of the disk on timescales of a year or two.  \cite{gal01} measured a correlation between the variability of redshifted and blueshifted maser features in NGC 1068, which they used to argue that the masers respond to variability in the central engine.

Since the fiducial picture of circumnuclear megamaser disk geometry (for a Keplerian rotation curve) allows us to uniquely associate any high-velocity maser feature with a radial location within the accretion disk, we attempt here to detect the propagation of some signal through the masing disks of our best-sampled targets.  A measurement of disk reverberation would not only lend support to the idea of a common pumping source, but it could also potentially enable an independent means of measuring the mass of the central SMBH and the distance to the host galaxy (provided the propagation velocity of the reverberation signal is known).  If we denote the outward propagation speed as $v_s$, then a reverberation signal passing through the spectrum at a rate $\dot{v}$ corresponds to a black hole mass of

\begin{equation}
M_{\text{BH}} = - \frac{v_s (v - v_0)^3}{2 G \dot{v}} . \label{eqn:ReverberationMass}
\end{equation}

\noindent Here, $v$ is the observed velocity (i.e., as seen in the spectrum) and $v_0$ is the velocity of the dynamic center (i.e., the motion of the black hole itself, which is presumably almost identical to the recession velocity of the galaxy).  We note that $\dot{v}$ will in general be a function of $v$; that is, for a constant value of $v_s$ the rate at which the reverberation signal passes through the spectrum depends on where in the spectrum it is located.  Once the black hole mass is known, the distance to the galaxy can be determined by comparing the angular orbital radii of the maser spots (measured using VLBI) to the orbital radii calculated using the single-dish spectra (from $r = G M_{\text{BH}} / v^2$).

\subsection{Extracting a reverberation signal} \label{ReverberationAnalysis}

\begin{figure*}
	\centering
		\includegraphics[width=1.00\textwidth]{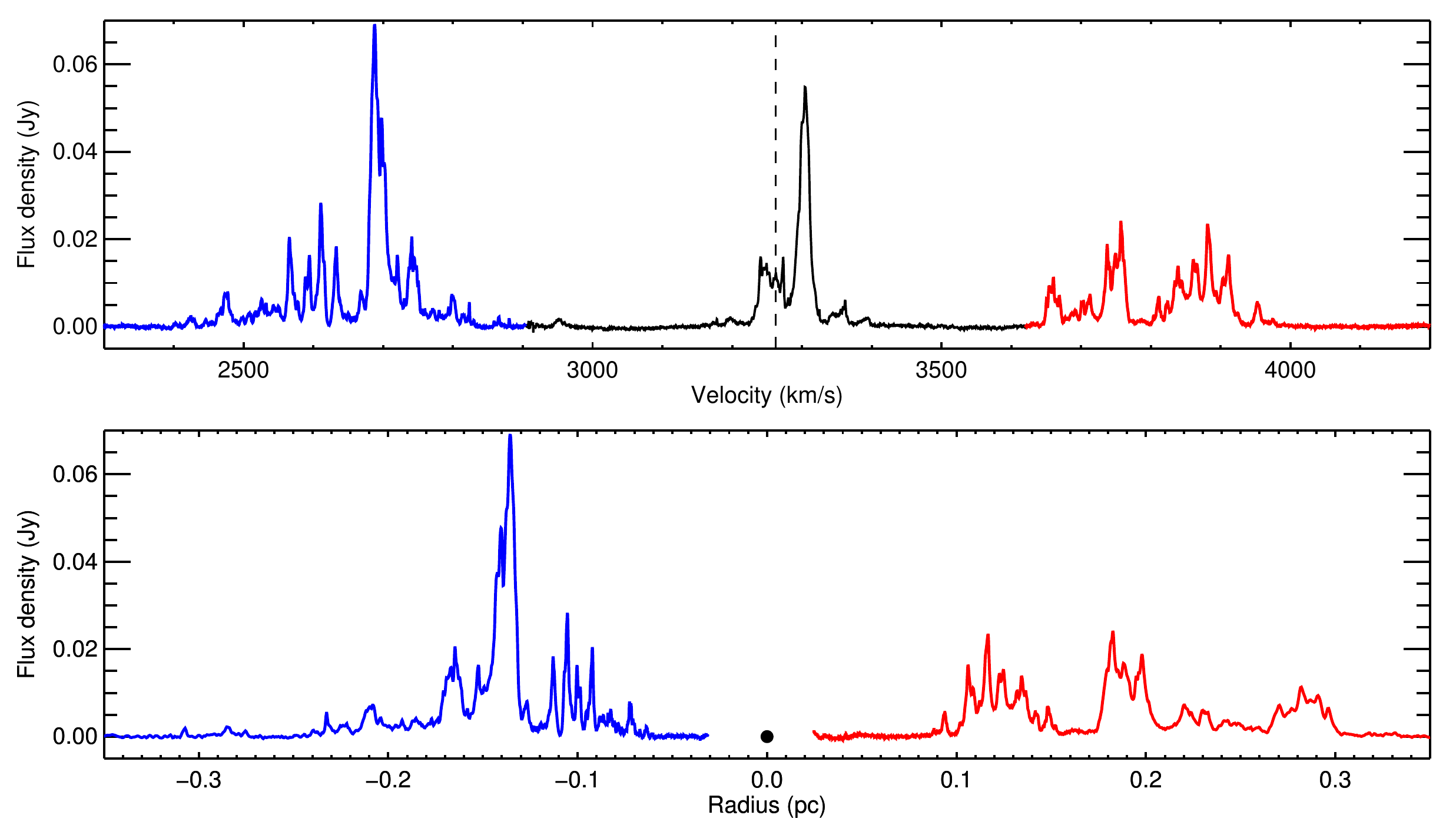}
	\caption{Illustration of the conversion between a velocity spectrum (top) and a radial spectrum (bottom), using Equation \ref{eqn:SpectrumConversion}.  The dashed line in the upper spectrum shows the recession velocity of the system, and the black point in the lower spectrum shows the location of the SMBH.  The blueshifted portion of each spectrum is plotted in blue, while the redshifted portion is plotted in red.  The source chosen for this example is the galaxy UGC 3789.}
	\label{fig:spectrum_conversion}
\end{figure*}

Here we outline the procedure used to check for the spectral signature of radially-propagating excitation in a time series of GBT disk maser spectra.  The relevant parameters are the mass of the central black hole, $M_{\text{BH}}$, the recession velocity of the dynamic center, $v_0$, and the propagation speed of the signal, $v_s$.  The observed response of a high-velocity maser offset by a distance $D$ (see bottom panel in Figure \ref{fig:spectrum_conversion}) is delayed by $D/v_s$ relative to the response of all systemic masers.

We subtract a weighted average spectrum (see \S\ref{DiskSpectra}) of the target from each epoch to remove stable (i.e., non-propagating) high-velocity features from each spectrum.  We then map each velocity channel, $v_i$, to a radial position, $r_i$, within the maser disk.  The mapping assumes that the high-velocity maser spots are all located on the midline of the disk, and that they are all on circular Keplerian orbits:

\begin{equation}
r_i = \frac{G M_{\text{BH}}}{(v_i - v_0)^2} . \label{eqn:SpectrumConversion}
\end{equation}

\noindent We refer to the original GBT spectra as the ``velocity spectra" and the new, radially-mapped spectra as the ``radial spectra."  An example of these two for the source UGC 3789 is shown in Figure \ref{fig:spectrum_conversion}.

\begin{figure*}
	\centering
		\includegraphics[width=1.00\textwidth]{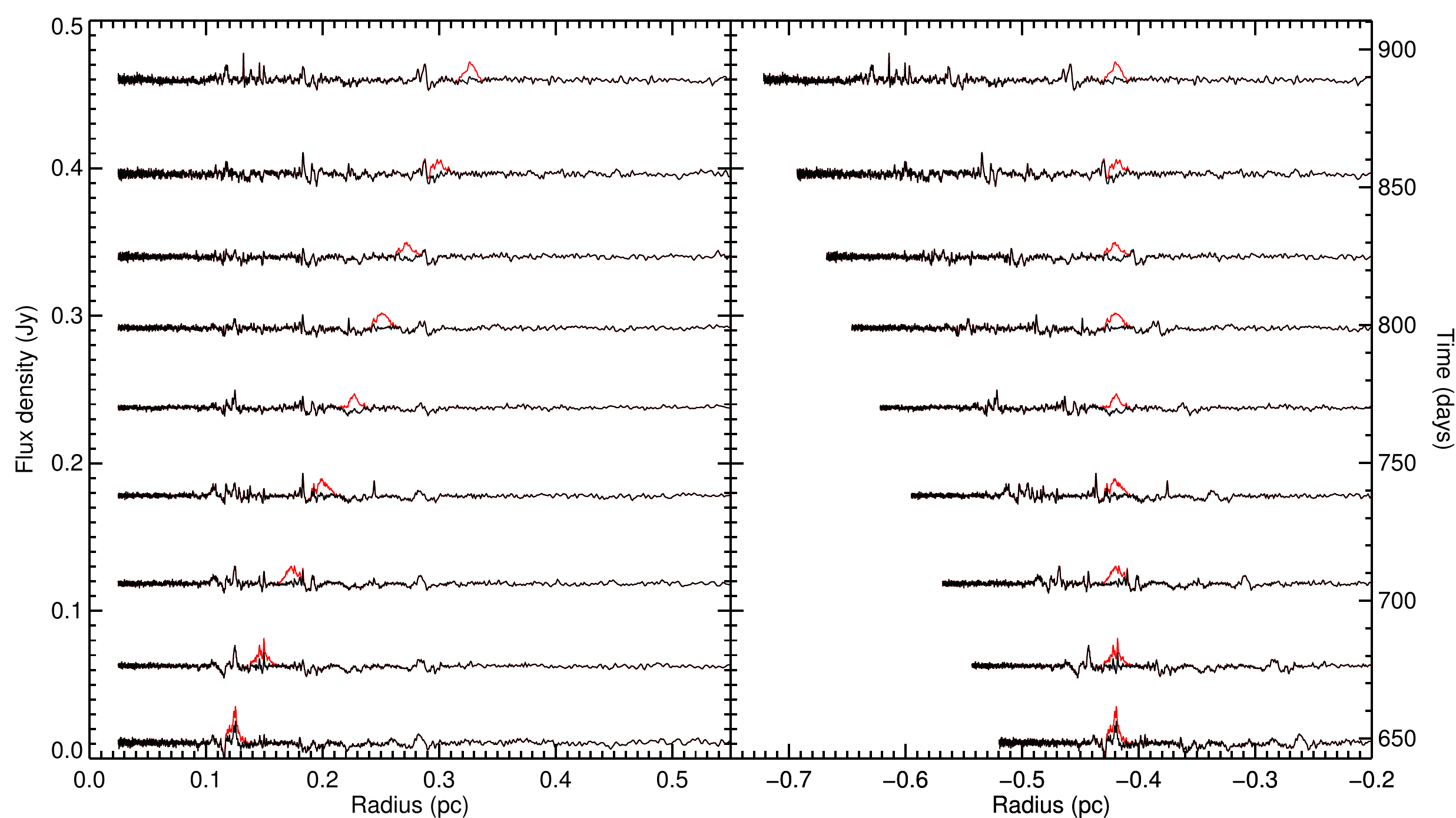}
	\caption{These plots show the radial spectra before (left) and after (right) accounting for the time delay caused by the propagation of the signal.  The black spectra are real spectra of UGC 3789, and the red line includes the artificially injected 10 mJy signal.  In the panel on the left, we can see that the artificial signal is propagating outwards with time.  In the panel on the right, the spectra have been temporally shifted using $v_s = c$; as a result, when stacking these spectra the signal will add coherently.  In both panels, the spectra have been vertically offset by an amount proportional to the time between observations; the time since the first observation is shown on the right axis.  The radial zeropoint corresponds to the position of the SMBH at the time of the first observation.  Only the redshifted high-velocity features are shown in these plots.}
	\label{fig:stacking_signal}
\end{figure*}

To account for the time delay between the detection of a propagating signal in consecutive epochs, each radial spectrum is temporally shifted according to the signal propagation speed and that spectrum's date of observation, relative to some reference epoch.  For simplicity, we have defined the temporal zeropoint to be the date of the first observation, given in Figure \ref{fig:AveragedSpectra}.  This process is illustrated in Figure \ref{fig:stacking_signal}.

After shifting, the radial spectra are then averaged over all epochs.  If a target has been observed for $N$ epochs, each of which has an associated radial spectrum $S_n(r,t_n)$, then this procedure can be written as

\begin{equation}
S(r) = \frac{1}{N} \sum_{n=1}^N S_n(r - v_s t_n , t_n) .
\end{equation}

\noindent Here, $S(r)$ is the final combined radial spectrum.  The radial zeropoint is defined to be the center of the disk (i.e., the location of the SMBH) at the date of the first observation.

The purpose of this procedure is to stack spectra in such a way that a radially-propagating signal will add coherently across all epochs.

\subsection{Sensitivity} \label{ReverbMapSensitivity}

The sensitivity of our method depends on several factors, including the number of epochs and overall time baseline of observation, as well as the intrinsic variability of the target.  We restrict our analysis to well-sampled (i.e., $\gtrsim$20 epochs of observation) galaxies that have reliably measured black hole masses \citep[see][]{kuo11}.

Some of these targets are more variable than others.  In general, the more flaring a source displays, the less sensitive this measurement will be.  Flaring events are not removed well when subtracting an epoch-averaged spectrum, and so sufficiently strong flares can appear as false positives in the final radial spectrum.

Although we've chosen to test only those sources for which $M_{\text{BH}}$ is known to $\sim$10\% or better, it's possible that our method requires the value to be even more precisely known to ensure recovery of a propagating signal.  To test the sensitivity of our method on the input values of $v_s$ and $M_{\text{BH}}$, we injected an artificial propagating signal into a series of spectra.  The signal was a Gaussian pulse of fixed width and amplitude, propagating with a fixed velocity from a black hole of known mass.  We found that the tolerance threshold for both $v_s$ and $M_{\text{BH}}$ was approximately 5\%; if either of these inputs is off from the true value by more than this amount, the signal is not recovered or is severely degraded.

\subsection{Discussion} \label{ReverberationDiscussion}

We tested for reverberation in the six maser galaxies listed in Table \ref{tab:Reverb_map_targets}.  For each galaxy, we checked for signals propagating at velocity increments of $0.01 c$, with minimum and maximum propagation velocities of $0.8 c$ and $1.2 c$, respectively\footnote{We actually investigated propagation speeds down to $0.0 c$, but the sensitivity of the method starts to drop considerably below a certain speed.  This is because the individual maser features -- which in general aren't perfectly matched to the average spectrum, so they don't subtract out well -- begin to add semi-coherently, rather than averaging out like noise.  To give an example, the 3$\sigma$ threshold for UGC 3789, which is about 0.8 mJy for propagation speeds between $0.8 c$ and $1.2 c$, increases to $\sim$5 mJy for a propagation speed of $0.5 c$.  This also makes it more difficult to differentiate between a propagating signal and a coherently-added maser feature, so we only quote sensitivities between $0.8 c$ and $1.2 c$ for Table \ref{tab:Reverb_map_targets}.}.  We also adjusted the black hole masses within a range $\pm 20$\% of the measured value, in increments of 1\%.  No reverberation signals were detected in any of the galaxies, with limiting flux densities listed in Table \ref{tab:Reverb_map_targets}.  Given that the spectra for these galaxies typically vary at the $\sim$tens of mJy level (see \S\ref{Variability}), we can see that any contribution from a propagating signal must constitute only a small ($\lesssim$10\%) fraction of the total variability.

The detection thresholds listed in Table \ref{tab:Reverb_map_targets} are simply the 3$\sigma$ noise levels in the final combined spectra.  We emphasize that this threshold gives only a limiting value for a signal that is perpetually coherent (i.e., always maintains its profile shape and moves at constant velocity) and that is present in all available spectra (i.e., it does not fade in and out as it propagates).  This procedure is less sensitive to a more complex signal.

Furthermore, we note that the timescale for variability of the pumping source influences our measurements.  If the source doesn't vary much over the $\sim$few-year timescales probed by these data, then the signal won't be radially localized and our technique will not help to detect it.

\begin{deluxetable}{lcccc}
\tablecolumns{4}
\tablewidth{0pt}
\tablecaption{\label{tab:Reverb_map_targets}}
\tablehead{Target &	\colhead{$v_0$ (km s$^{-1}$)}	&	\colhead{$M_{\text{BH}}$ ($10^7 \text{ M}_{\odot}$)} & \colhead{Epochs} & \colhead{Threshold (mJy)}}
\startdata
UGC 3789			&	3262	&	1.04	&	58	&	0.8	\\
Mrk 1419			&	4954	&	1.16	&	55	&	1.4	\\
NGC 6323			&	7829	&	0.94	&	44	&	1.2	\\
NGC 1194			&	4063	&	6.5		&	43	&	4.1	\\
NGC 2273			&	1832	&	0.75	&	38	&	1.8	\\
NGC 6264			&	10194	&	2.91	&	28	&	0.8	\\
\enddata
\tablecomments{Galaxies tested for a reverberation signal.  The threshold column lists the $3 \sigma$ detection cutoffs; a signal stronger than this value would be classified as a detection.  Note that the velocity of the dynamic center ($v_0$) need not be the same as the recession velocity of the galaxy listed in Table \ref{tab:Maoz_McKee_targets}, as the $v_0$ values were obtained by fitting Keplerian rotation curves to position-velocity data \citep{kuo11}.}
\end{deluxetable}

\section{Magnetic field strengths from Zeeman splitting} \label{MagneticFieldLimits}

Magnetic fields in AGN accretion disks are thought to drive several important physical processes.  The magnetorotational instability (MRI), first described in a general astrophysical context by \cite{bal91}, is likely the primary means by which angular momentum is transported in accretion disks.  Magnetic fields are also necessary for launching outflows, from the classic MHD disk wind \citep{bla82} to more modern incarnations that also incorporate radiation pressure (e.g., \citealt{kea12}).  In this section, we use measurements of the Zeeman effect to place limits on the magnetic field strength in several megamaser disks.

The maser emission that we observe at 22 GHz arises from one or more of the six hyperfine transitions of the $6_{16}$-$5_{23}$ rotational transition of the water molecule \citep[see][]{fie89}.  Since this molecule is non-paramagnetic, Zeeman splitting of these hyperfine energy levels arises from the coupling between the nuclear magnetic moments and an external magnetic field.  This causes the effect to be much weaker (by a factor of $\sim$$10^3$) in water than in molecules such as OH, where the unpaired electron's spin couples with the magnetic field.  The drastic difference in magnitude arises because the Bohr magneton and the nuclear magneton differ by the ratio of the electron to the nucleon mass, $m_e / m_p \approx 1/1836$ \citep{hei93}.

An external magnetic field causes each hyperfine level to split into three groups of lines: the $\pi$ components and the $\sigma^{\pm}$ components, corresponding to magnetic quantum number changes of $\Delta M_F = 0$ and $\Delta M_F = \pm 1$, respectively \citep{mod05}.  The $\sigma^{\pm}$ components are circularly polarized about the magnetic field direction, and they are symmetrically offset from the parent frequency.  For weak magnetic fields (i.e., $B \lesssim 1$ Gauss), this frequency offset is small compared to the line width; typically $(\Delta v_z / \Delta v_L) \sim 10^{-3}$--$10^{-4}$.

\subsection{Method} \label{ZeemanMethod}

In principle, the measured frequency difference between the left and right circular polarizations (corresponding to $\sigma^+$ and $\sigma^-$, respectively) allows us to determine the line-of-sight component of the magnetic field at the location of the maser spot.  Since the offset is small compared to the width of the line profile, the Stokes $V$ profile (given by $V = [\text{LCP} - \text{RCP}]/2$) is proportional to the derivative of the Stokes $I$ profile (given by $I = [\text{LCP} + \text{RCP}]/2$).  This leads to a characteristic S-shape of the Stokes $V$ profile (see, e.g., \citealt{vle01}, Fig. 2).

\cite{mod05} conducted a series of Monte Carlo simulations which established that the RMS sensitivity to the line-of-sight component of the magnetic field from a single maser line is consistent with what one would expect from a statistical treatment (see, e.g., \citealt{len92}), namely:

\begin{equation}
\sigma_B = \frac{\Delta v_L}{2 A} \left[ \frac{S}{N} \right]^{-1} . \label{eqn:ZeemanSensitivity}
\end{equation}

\noindent Here, $\Delta v_L$ is the FWHM line width, $S/N$ is the Stokes $I$ signal-to-noise ratio, and $A$ is the Zeeman splitting coefficient (which is different for each hyperfine transition).  After numerically solving the radiative transfer and rate equations for magnetized water masers, \cite{ned92} found that a value for $A$ of 0.020 km s$^{-1}$ G$^{-1}$ was most appropriate for the merging of the three dominant hyperfine components.  This value assumes that the three strongest hyperfine lines all contribute to a given observed maser line, and deviations from this value never exceeded a factor of $\sim$2 across the range of parameter space investigated in \cite{ned92}.  We thus adopt $A = 0.02$ km s$^{-1}$ G$^{-1}$ for our calculations, which in general follow the same procedure outlined in \cite{mod05}.

For extragalactic sources, only three efforts to measure magnetic field strengths using the Zeeman effect in H$_2$O megamasers have been published.  \cite{mod05} placed a 1$\sigma$ upper limit of 30 mG on the radial component of the magnetic field in NGC 4258, using a cross-correlation method to handle the heavy blending of the spectral features.  \cite{vle07} used the same technique on NGC 3079, obtaining an upper limit of 11 mG for the blueshifted features.  Both studies also measured limits for strong, isolated maser components, and combined these results with those from the cross-correlation method.  Additionally, \cite{mcc07} measured isolated lines to place a 1$\sigma$ upper limit of 50 mG on the toroidal component of the magnetic field in the Circinus galaxy.

\subsection{Measurements} \label{ZeemanMeasurements}

The most sensitive test for Zeeman splitting using individual (i.e., non-blended) maser lines occurs on lines that are both strong (large signal-to-noise) and narrow (small $\Delta v_L$).  We therefore focused our test on strong ($S/N > 50$) flaring events.

For each selected maser flare, we separately reduced the LCP and RCP spectra without applying Hanning smoothing; this process retains the full spectral resolution.  To compensate for errors in flux scale calibration, the peak value of the RCP spectrum was scaled to the value of the LCP spectrum prior to computing either the Stokes $I$ or Stokes $V$ spectra.  Typical scaling offsets were of order 10\%.  We note that the absolute intensity scale is unimportant for these measurements.

\begin{figure*}[p]
	\includegraphics[width=1.00\textwidth]{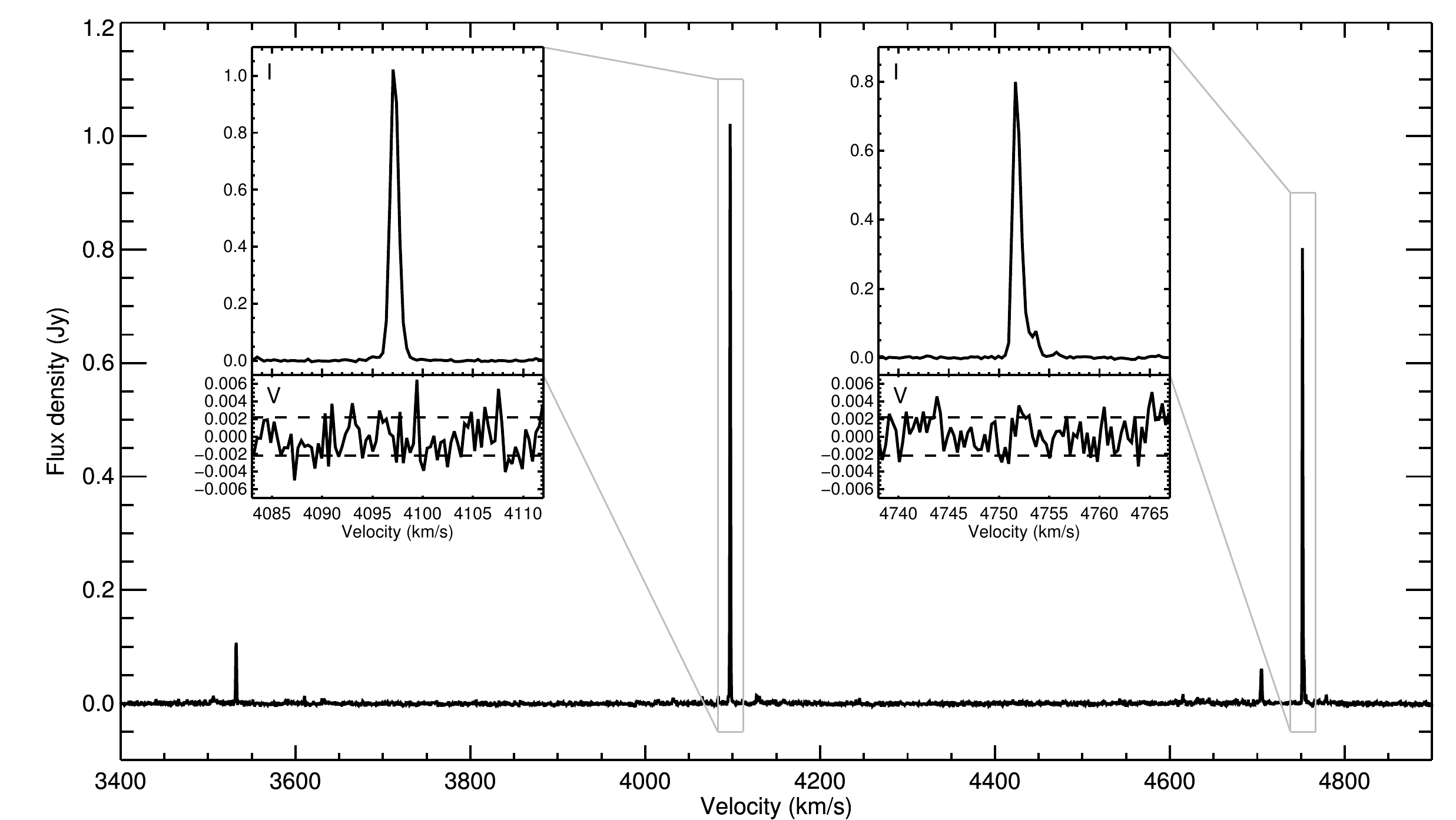}
	\caption{GBT spectrum of NGC 1194, taken on 2011 December 30.  Inset are the Stokes $I$ and $V$ profiles for the 4097.2 km s$^{-1}$ feature (left) and the 4751.7 km s$^{-1}$ feature (right).  The black dashed lines in the Stokes $V$ plots show the $1 \sigma$ RMS level for this spectrum.  No Zeeman profile is evident for either of these lines; limits are given in Table \ref{tab:Zeeman_targets}.}
	\label{fig:zeeman_NGC1194}
\end{figure*}

We did not detect Zeeman splitting in any of the maser lines, so our results here yield only upper limits on the magnetic field strengths.  These results are summarized in Table \ref{tab:Zeeman_targets}, and an example measurement (from NGC 1194) is shown in Figure \ref{fig:zeeman_NGC1194}.

\begin{deluxetable}{lccccccccc}
\tabletypesize{\footnotesize}
\tablecolumns{10}
\tablewidth{0pt}
\tablecaption{\label{tab:Zeeman_targets}}
\tablehead{	&		&	\colhead{$M_{\text{BH}}$}	&	\colhead{$v$}	&	\colhead{$V_{\text{rot}}$}	&	\colhead{Peak}	&	\colhead{$S/N$}	&	\colhead{$\Delta v_L$}	& \colhead{$B_{\parallel}$}	&	\colhead{Radius} \\
Target	&	\colhead{Date}	&	\colhead{($10^7$ M$_{\odot}$)}	&	\colhead{(km s$^{-1}$)}	&	\colhead{(km s$^{-1}$)}	&	\colhead{(mJy)}	&		&	\colhead{(km s$^{-1}$)}	&	\colhead{(mG)}	&	\colhead{(pc)}}
\startdata
NGC 1194			&	2007 Dec 26							&	6.5$^a$		&	$4757.6$	&	$694.6$		&	340		&	141	&	0.58	&	$<$100 (t)	&	0.58	\\
NGC 1194			&	2010 Apr 10							&	6.5				&	$4146.4$	&	$83.4$		&	210		&	97	&	0.87	&	$<$220 (r)	&	-			\\
NGC 1194			&	2011 Dec 30							&	6.5				&	$4097.2$	&	$34.2$		&	1020	&	330	&	0.96	&	$<$73 (r)		&	-			\\
NGC 1194			&	2011 Dec 30							&	6.5				&	$4751.7$	&	$688.7$		&	800		&	259	&	0.95	&	$<$91 (t)		&	0.59	\\
NGC 2273			& 2009 Dec 12							&	0.75$^a$	&	$1582.5$	&	$-249.5$	&	240		&	115	&	0.72	&	$<$160 (t)	&	0.52	\\
NGC 3393			&	2006 Apr 28							&	3.1$^b$		&	$4050.9$	&	$300.9$		&	230		&	95	&	0.84	& $<$220 (t)	&	1.48	\\
NGC 3393			&	2006 Dec 6\hphantom{0}	&	3.1				&	$4260.8$	&	$510.8$		&	350		&	94	&	1.1		& $<$300 (t)	&	0.51	\\
UGC 3789			&	2010 Dec 20							&	1.04$^a$	&	$3273.0$	&	$11.0$		&	190		&	75	&	0.74	&	$<$250 (r)	&	-			\\
NGC 6323			&	2008 Mar 25							&	0.94$^a$	&	$7395.2$	&	$-433.8$	&	180		&	86	&	1.0		&	$<$300 (t)	&	0.21	\\
ESO 558-G009	&	2013 Apr 22							&	1.8$^c$		&	$8003.7$	&	$329.7$		&	490		&	81	&	0.99	&	$<$310 (t)	&	0.71	\\
Mrk 1419			&	2007 Apr 14							&	1.16$^a$	&	$5330.8$	&	$376.8$		&	220		&	56	&	1.6		&	$<$720 (t)	&	0.35	\\
\enddata
\tablecomments{Maser lines tested for Zeeman splitting.  For flaring lines appearing in more than one epoch, the listed observation date is that which yields the best upper limit on the line-of-sight component of the magnetic field.  In addition to the Doppler velocity ($v$), we list the rotation velocity ($V_{\text{rot}} = v - v_0$; blueshifted lines are negative, $v_0$ is the velocity of the dynamic center) and the measured line width ($\Delta v_L$) for each line.  For all lines, $B_{\parallel}$ is quoted as a $1 \sigma$ upper limit, and the letters in parentheses indicate whether the measurement is sensitive to the toroidal (t) component or the radial (r) component.  For limits on toroidal magnetic field components, the radius column gives the corresponding radial location in the disk at which the limit holds. \\ \vspace{2mm}
$^a$\cite{kuo11} \\
$^b$\cite{kon08} \\
$^c$Gao et al. (\textit{submitted})}
\end{deluxetable}

\subsection{Discussion} \label{ZeemanDiscussion}

Since the Zeeman measurements are only sensitive to the line-of-sight magnetic field, $B_{\parallel}$, the high-velocity and systemic lines measure different equatorial components of this field.  The high-velocity features measure the toroidal component of the field, $B_{\text{tor}}$, while the systemic features measure the radial component, $B_\text{r}$.  None of the features directly measure the poloidal component of the magnetic field, but an appropriate model (see, e.g., \citealt{haw96}) can estimate its magnitude using the values of the toroidal and radial components.

Even without knowledge of the poloidal component, we can still use the derived upper limits to constrain the support mechanism for the accretion disks.  This is because only the components of the field that thread through the disk (i.e., only the radial and toroidal components) can provide vertical pressure support.  For typical maser conditions of $n \approx 10^9$ cm$^{-3}$ and $T \approx 1000$ K, the gas pressure amounts to roughly $10^{-4}$ erg cm$^{-3}$.  The equivalent support from magnetic pressure would require a $\sim$50 mG magnetic field, which is comparable to (though still slightly below) our most stringent limits.  It is worth noting that these numbers are also comparable to the $\sim$100 mG upper limit imposed by hydrostatic equilibrium for the disk thickness measured by \cite{arg07} in NGC 4258.

\section{Summary} \label{Summary}

We have addressed several new scientific questions that can be explored using the MCP's extensive monitoring campaign of 22 GHz disk megamaser spectra with the GBT.  The spectra in this dataset are unique in their ability to probe the accretion disks of nearby AGN at sub-parsec scales, and the dataset itself is unmatched in the sensitivity and time coverage for each target.

\begin{enumerate}
	\item We present a comprehensive collection of Keplerian disk megamaser spectra.  We also present dynamic spectra for the most heavily monitored of these sources.
	\item We find that the redshifted high-velocity maser features are brighter, on average, than the blueshifted features for our sample of 32 megamaser disks.  This asymmetry is predicted by the spiral shock model of MM98.
	\item We also test the MM98 prediction that the high-velocity features should exhibit nonzero line-of-sight velocity drifts.  We find no systematic drifts.  Furthermore, the statistically significant detection of both positive and negative velocity drifts within the same set of features (as we have for several sources) is inconsistent with the MM98 model's predictions.
	\item We argue that the intra-day variability observed in ESO 558-G009 is most likely caused by ISS, and we derive parameters of the scattering screen under different assumptions about the scattering regime.  Though the measurements are currently sparse, we find that they are most consistent with a relatively nearby ($\sim$70 pc) scattering screen.
	\item We test six maser systems for a radially-propagating change in maser activity, which could be the result of variable output from the central engine.  No such signal is detected in any of the galaxies.
	\item We measure upper limits on the toroidal and radial magnetic field strengths in the accretion disks of 7 galaxies using the Zeeman effect, and we find that the magnetic fields must be less than several hundred mG in each case.  This is beginning to probe the regime where the magnetic pressure becomes comparable to the gas pressure in the disk.
\end{enumerate}

\acknowledgments

We acknowledge Fred Schwab for guidance on the statistical analysis, Ken Kellerman for a discussion of variability timescales, and Scott Suriano for helpful clarifications regarding disk wind phenomenology.  The National Radio Astronomy Observatory is a facility of the National Science Foundation operated under cooperative agreement by Associated Universities, Inc.  This research was supported in part by an ARCS/MWC Scholar Award.  This research has made use of the NASA/IPAC Extragalactic Database (NED), which is operated by the Jet Propulsion Laboratory, California Institute of Technology, under contract with the National Aeronautics and Space Administration.

{\it Facilities:} \facility{GBT}

\bibliographystyle{apj}
\bibliography{mybib}

\appendix

\section{Kernel density estimation (KDE)} \label{app:KDE}

In essence, the KDE technique as used in this paper is simply an alternative to a traditional histogram (though in each case we have shown it alongside such a histogram).  In a standard histogram, a single data point falls into a ``bin" of width $h$, unit height, and fixed edgepoints.  The bin width is usually determined by the sample size and spread, with the optimal result being a compromise between data resolution and population per bin.  The bin edgepoints, however, are often more arbitrarily defined.  The KDE approach solves this issue by eliminating the use of bins; instead, each data point is represented by a ``kernel" of some predefined functional form.  In Figures \ref{fig:histogram_luminosities} and \ref{fig:histogram_MaozMcKee} we used a Gaussian kernel of the form

\begin{equation}
K(u) = \frac{1}{\sqrt{2 \pi}} e^{-u^2/2}. \label{eqn:KDEkernel}
\end{equation}

\noindent This kernel has been scaled by $h$ relative to a normal (i.e., unit area) Gaussian kernel, such that the area under the curve for a given data point matches what would be found in a typical histogram of bin width $h$.  The final kernel density estimator is then just a sum of the kernels for all data points, which can be written as

\begin{equation}
f(x) = \sum_{i=1}^N K\left( \frac{x - X_i}{h} \right). \label{eqn:KDEfunction}
\end{equation}

\noindent Here, $N$ is the number of data points, $X_i$ is the center of the kernel for data point $i$ (i.e., the value of that data point), and $h$ is the kernel width.  In our case, $X_i$ is the isotropic luminosity (for Figure \ref{fig:histogram_luminosities}) or the logarithm of the flux ratio (for Figure \ref{fig:histogram_MaozMcKee}) for a single source.  Under the assumption that our underlying distribution is at least approximately Gaussian, we have used the bin width derived by \citeauthor{sil86} (page 45, equation 3.28):

\begin{equation}
h = \left( \frac{4 \sigma^5}{3 N} \right)^{1/5}. \label{eqn:KDEbandwidth}
\end{equation}

\noindent Here, $\sigma$ is the standard deviation of the sample.

\section{Transverse motion along the line of sight to ESO 558-G009} \label{app:TransverseMotion}

Our goal is to transform from the Galactic Cartesian coordinate system $(X,Y,Z)$ to a coordinate system $(x,y,z)$ where the line of sight to ESO 558-G009 is aligned with the $z$-axis.  We define $\hat{\boldsymbol{z}}$ to be pointing away from ESO 558-G009 and $\hat{\boldsymbol{y}}$ to be the projection of the North Ecliptic Pole onto the plane perpendicular to $\hat{\boldsymbol{z}}$.  The unit vector $\hat{\boldsymbol{x}}$ is then defined to be $\hat{\boldsymbol{x}} \equiv \hat{\boldsymbol{z}} \times \hat{\boldsymbol{y}}$.

The standard spherical Galactic coordinates $(\ell,b)$ can be converted to Galactic Cartesian unit vectors using the transformation:

\begin{align}
X &= \cos(\ell) \cos(b) \\ \nonumber
Y &= \sin(\ell) \cos(b) \\ \nonumber
Z &= \sin(b)
\end{align}

\noindent We can thus define a unit vector $\hat{\boldsymbol{r}} = (X,Y,Z)$ that points in the direction of any Galactic coordinate location $(\ell,b)$.

The North Ecliptic Pole has Galactic coordinates $(\ell,b) = (96.3840,29.8117)$, with corresponding unit vector $\hat{\boldsymbol{r}}_{\text{NEP}} = (-0.0965,0.8623,0.4972)$.  The coordinates for ESO 558-G009 are $(\ell,b) = (233.6609,-6.9598)$, with unit vector $\hat{\boldsymbol{r}}_{\text{ESO}} = (-0.5882,-0.7996,-0.1212)$.  From our description above of the desired coordinate system, we have the following expressions for the coordinate unit vectors:

\begin{align}
\hat{\boldsymbol{x}} &= \frac{\hat{\boldsymbol{r}}_{\text{NEP}} \times \hat{\boldsymbol{r}}_{\text{ESO}}}{\left| \hat{\boldsymbol{r}}_{\text{NEP}} \times \hat{\boldsymbol{r}}_{\text{ESO}} \right|} \\ \nonumber
\hat{\boldsymbol{y}} &= \hat{\boldsymbol{r}}_{\text{ESO}} \times \hat{\boldsymbol{x}} \\ \nonumber
\hat{\boldsymbol{z}} &= - \hat{\boldsymbol{r}}_{\text{ESO}}
\end{align}

\noindent These evaluate to $\hat{\boldsymbol{x}} = (0.4064,-0.4218,0.8105)$, $\hat{\boldsymbol{y}} = (-0.6992,0.4275,0.5731)$, and $\hat{\boldsymbol{z}} = (0.5882,0.7996,0.1212)$.  We'll henceforth refer to this new coordinate system as the ``source" coordinate system.

\subsection{Solar motion with respect to the LSR} \label{app:SolarMotionLSR}

The first component of the transverse motion comes from the Sun's deviation from its orbital motion.  From \cite{cos11}, the Sun's peculiar motion relative to the LSR has components $\boldsymbol{v}_{\odot} = (8.50,13.38,6.49)$ km s$^{-1}$, with magnitude $v_{\odot} = 17.13$ km s$^{-1}$ and corresponding unit vector $\hat{\boldsymbol{r}}_{\odot} = (0.4962,0.7811,0.3789)$.  The parallel and perpendicular components of this velocity are then simply its projections onto the coordinate axes:

\begin{align} \label{eqn:SolarLSR}
v_{\odot,\parallel} &= \boldsymbol{v}_{\odot} \cdot \hat{\boldsymbol{z}} \\ \nonumber
v_{\odot,\perp} &= \sqrt{\left( \boldsymbol{v}_{\odot} \cdot \hat{\boldsymbol{x}} \right)^2 + \left( \boldsymbol{v}_{\odot} \cdot \hat{\boldsymbol{y}} \right)^2}
\end{align}

\noindent These evaluate to $v_{\odot,\parallel} = 16.48$ km s$^{-1}$ and $v_{\odot,\perp} = 4.66$ km s$^{-1}$, with source components $(v_x,v_y,v_z)_{\odot} = (3.07,3.50,16.48)$ km s$^{-1}$.

\subsection{Earth's orbital motion} \label{app:EarthMotion}

The second component of the transverse motion comes from the Earth's orbit around the Sun.  For simplicity, we'll model this orbit as circular about the North Ecliptic Pole, with orbital velocity $v_{\oplus} = 30$ km s$^{-1}$.  If we define a position angle $\phi = \omega t$ measured clockwise from the negative $x$-axis, then we can decompose the Earth's orbital motion into the source components:

\begin{align} \label{eqn:EarthOrbit}
v_{x,\oplus}(t) &= v_{\oplus} \sin(\phi_0 + \omega t) \\ \nonumber
v_{y,\oplus}(t) &= v_{\oplus} \cos(\phi_0 + \omega t) \cos(i) \\ \nonumber
v_{z,\oplus}(t) &= -v_{\oplus} \cos(\phi_0 + \omega t) \sin(i)
\end{align}

\noindent Here, $\omega$ is the orbital angular frequency of the Earth, $i = \pi/2 - \cos^{-1}\left( \hat{\boldsymbol{y}} \cdot \hat{\boldsymbol{r}}_{\text{NEP}} \right)$ is the inclination of the orbit relative to the line of sight to ESO 558-G009 (in our case, $i = 46.1^{\circ}$), and $\phi_0$ is an initial position angle that must be calibrated based on the known motion of the Earth.

On the vernal equinox (the origin of the ecliptic longitude), the Earth is moving towards ecliptic coordinates $(\lambda,\beta) = (90,0)$.  The equivalent Galactic coordinates are $(\ell,b) = (186.3725,-0.0200)$, so the corresponding velocity vector is $\boldsymbol{v}_{\oplus,\text{eq}} = (-29.814,-3.33,0.009)$ km s$^{-1}$.  Decomposing this into source coordinates yields $(v_x,v_y,v_z)_{\oplus,\text{eq}} = (-10.704,19.428,-20.199)$ km s$^{-1}$.

Since our model uses only a crude approximation for what in reality is a moderately noncircular orbit, small deviations from the model will grow with time.  As such, we'd like to calibrate it using the vernal equinox closest in time to the observations.  This occurred on 2012 May 20, which corresponds to a Modified Julian Date of $\text{MJD} = 56006$.  We obtain a value of $\phi_0 = 6.096$.

\subsection{Solar orbital motion} \label{app:SolarMotionOrbital}

The third component of the transverse motion comes from the Sun's orbit about the Galactic center, relative to that of the scattering screen.  From \cite{rei14}, the distance from the Galactic center to the Sun is $R_0 = 8.34$ kpc.  If we denote the distance from the Sun to the scattering screen as $D$ and the distance from the scattering screen to the Galactic center as $R$, then the law of cosines gives us an expression:

\begin{equation}
R = \sqrt{D^2 + R_0^2 + 2 D R_0 \cos(\theta)} \label{eqn:FirstSolarR}
\end{equation}

\noindent Here, $\theta$ is the angle between $\ell = 180^{\circ}$ and the direction to ESO 558-G009 (i.e., $\theta = \ell - 180^{\circ}$).

If we define $\alpha$ to be the angle between the Sun and the scattering screen, as seen from the Galactic center, then we have a second expression for $R$:

\begin{equation}
R = D \cos(\theta - \alpha) + R_0 \cos(\alpha) \label{eqn:SecondSolarR}
\end{equation}

\noindent Combining Equations \ref{eqn:FirstSolarR} and \ref{eqn:SecondSolarR} yields a numerically invertible expression for $\alpha$ in terms of $D$.  Once we know $\alpha$, we can use it to determine the component of the scattering screen's orbital motion that lies along the same direction as the Sun's orbital motion.  If the orbital velocity of the scattering screen is $V_s$, then the parallel component is just $V_s \cos(\alpha)$.

The line of sight towards ESO 558-G009 is such that the scattering screen lies outside of the solar orbit.  The rotation curve of the Milky Way is known to be very nearly flat at these outer radii (see \citealt{rei14}), with an orbital velocity of 240 km s$^{-1}$.  We can thus set $V_s = V_{\odot} = 240$ km s$^{-1}$, and we obtain a net apparent motion of the scattering screen (directed along the Sun's orbital velocity vector) of:

\begin{equation}
V_{\parallel} = V_{\odot} \big( 1 - \cos(\alpha) \big)
\end{equation}

\noindent The Sun's orbital motion is directed towards the Galactic coordinates $(\ell,b) = (90,0)$, which is directed along the $Y$-axis.  The perpendicular component of the scattering screen's orbital velocity (i.e., the component directed along the $X$-axis) will then just be $V_{\perp} = - V_{\odot} \sin(\alpha)$.  We can now use our previously-derived unit vectors to transform this into the source frame.  Doing so yields:

\begin{align} \label{eqn:ScreenOrbit}
V_x &= V_{\odot} \Big[ - 0.4064 \sin(\alpha) - 0.4218 \big( 1 - \cos(\alpha) \big) \Big] \\ \nonumber
V_y &= V_{\odot} \Big[ 0.6992 \sin(\alpha) + 0.4275 \big( 1 - \cos(\alpha) \big) \Big] \\ \nonumber
V_z &= V_{\odot} \Big[ - 0.5882 \sin(\alpha) + 0.7996 \big( 1 - \cos(\alpha) \big) \Big] 
\end{align}

\noindent Combining this with Equations \ref{eqn:SolarLSR} and \ref{eqn:EarthOrbit} allows us to fully characterize the transverse motion of the scattering screen, as seen from Earth, in terms of $t$ (which is known for every observation) and $D$ (which we would like to know).  For a nearby screen, $D \ll R_0$, and the transverse motion becomes a function of $t$ only.

\end{document}